\documentclass{acmart}
\usepackage{algorithm}
\usepackage{algpseudocode}
\usepackage{listings}
\usepackage{subfig}
\usepackage{tikz} 
\usepackage{tabularx}
\usetikzlibrary{shapes,snakes}
\usepackage[inline]{enumitem}
\usepackage{multirow}
\definecolor{codegreen}{rgb}{0,0.6,0}
\definecolor{codegray}{rgb}{0.5,0.5,0.5}
\definecolor{codepurple}{rgb}{0.58,0,0.82}
\definecolor{backcolour}{rgb}{0.95,0.95,0.92}

\lstdefinestyle{mystyle}{
backgroundcolor=\color{backcolour},   
commentstyle=\color{codegreen},
keywordstyle=\color{magenta},
stringstyle=\color{codepurple},
basicstyle=\ttfamily\bfseries\footnotesize,
breakatwhitespace=false,         
breaklines=true,                 
captionpos=b,                    
keepspaces=true,                                     
numbersep=4pt,                  
showspaces=false,                
showstringspaces=false,
showtabs=false,                  
tabsize=1,
numbers=left,
xleftmargin=1.5em,
frame=single,
framexleftmargin=0.05em
}
\AtBeginDocument{%
  \providecommand\BibTeX{{%
    \normalfont B\kern-0.5em{\scshape i\kern-0.25em b}\kern-0.8em\TeX}}}

\setcopyright{acmcopyright}
\copyrightyear{2018}
\acmYear{2018}
\acmDOI{10.1145/1122445.1122456}

\acmConference[Woodstock '18]{Woodstock '18: ACM Symposium on Neural
  Gaze Detection}{June 03--05, 2018}{Woodstock, NY}
\acmBooktitle{Woodstock '18: ACM Symposium on Neural Gaze Detection,
  June 03--05, 2018, Woodstock, NY}
\acmPrice{15.00}
\acmISBN{978-1-4503-XXXX-X/18/06}

\begin{document}

\title{Machine Learning Enabled Scalable Performance Prediction of Scientific Codes}

\author{Gopinath Chennupati}
\orcid{0000-0002-6223-8570}
\affiliation{%
	\institution{Information Sciences (CCS-3) Group, Los Alamos National Laboratory}
	\streetaddress{Los Alamos}
	\city{Los Alamos}
	\country{NM}}
\email{gchennupati@lanl.gov}

\author{Nandakishore Santhi}
\affiliation{%
	\institution{Information Sciences (CCS-3) Group, Los Alamos National Laboratory}
	\streetaddress{Los Alamos}
	\city{Los Alamos}
	\country{NM}}
\email{nsanthi@lanl.gov}

\author{Phill Romero}
\affiliation{%
	\institution{HPC Environments, Los Alamos National Laboratory}
	\streetaddress{Los Alamos}
	\city{Los Alamos}
	\country{NM}}
\email{prr@lanl.gov}

\author{Stephan Eidenbenz}
\affiliation{%
	\institution{Information Sciences (CCS-3) Group, Los Alamos National Laboratory}
	\streetaddress{Los Alamos}
	\city{Los Alamos}
	\country{NM}}
\email{eidenben@lanl.gov}

\renewcommand{\shortauthors}{Chennupati et al.}

\begin{abstract}
	We present the Analytical Memory Model with Pipelines (AMMP) of the Performance Prediction Toolkit (PPT). PPT-AMMP takes high-level source code and hardware architecture parameters as input, predicts runtime of that code on the target hardware platform, which is defined in the input parameters. PPT-AMMP transforms the code to an (architecture-independent) intermediate representation, then (i) analyzes the basic block structure of the code, (ii) processes architecture-independent virtual memory access patterns that it uses to build memory reuse distance distribution models for each basic block, (iii) runs detailed basic-block level simulations to determine hardware pipeline usage. 
	
	PPT-AMMP uses machine learning and regression techniques to build the prediction models based on small instances of the input code, then integrates into a higher-order discrete-event simulation model of PPT running on Simian PDES engine. We validate PPT-AMMP on four standard computational physics benchmarks, finally present a use case of hardware parameter sensitivity analysis to identify bottleneck hardware resources on different code inputs. We further extend PPT-AMMP to predict the performance of scientific application (radiation transport), SNAP. We analyze the application of multi-variate regression models that accurately predict the reuse profiles and the basic block counts. The predicted runtimes of SNAP when compared to that of actual times are accurate.
\end{abstract}



\maketitle

\section{Introduction}
As traditional hardware scaling laws have started to fall apart, computer hardware designers have resorted to developing novel components on their chips that have changed the nature of hardware architecture usually with the goal of (i) exploiting parallelism at all levels of the computing stack and (ii) improving the performance of resource that pose bottlenecks. Modern hardware architectural features, such as parallel execution pipelines, vector units, speculative execution, branch prediction, threading, memory locality, memory hierarchies (4 or more levels deep), burst buffers, and fast interconnects have become standard equipment on most computing devices, from mobile phones to supercomputers. These advanced hardware features, and in particular their parallelization primitives, have to be matched with software stack that fully leverages these features. The software engineering industry has tried to keep up with these fast changes on the hardware side, with efforts such as (i) new runtime-systems for task-parallel execution (e.g., Legion \cite{legion}, Raja \cite{raja}, and Kokkos \cite{kokkos}) that serve as alternatives or extensions to the MPI model of parallel tasks, (ii) a renewed push towards multi-threading including OpenMP and OpenACC and (iii) many compiler optimization techniques aimed at increasing memory locality and improving pipeline usage.

In this emerging novel software/hardware ecotone, code performance (measured in runtime) can be non-linear and hard to estimate. Several attempts have built hardware-software performance modeling tools (see \cite{wsc2017kishwar} for a recent survey). Most of these tools focus on one/two layers of the hardware or software stack, greatly abstracting away other layers. There is a good reason for this: modeling and simulating a computer system is inherently a {\it multi-scale problem}, akin -- in this aspect only -- to the situation in computational physics, where some methods/tools excel at modeling subparticle physics, others compute at a molecular level, and yet others look at fluid dynamics of entire engineered or natural systems (such as global climate). Just as it is hopeless from a scalability perspective to simulate global climate by simulating every single atom of the more or less closed system ``earth'', we cannot expect to succeed at modeling each individual clock-cycle (at $10^{-9}$ second resolution) of a large supercomputer (with $10^7$ compute cores) with discrete-event simulation (which produces $10^{18}$ events for a simulated supercomputer run of one minute). Rather, we need to build a suite of models tackling different aspects of the computing stack with careful integration across dimension boundaries, including error estimates. 

In this paper, we describe the multi-scale performance modeling tool PPT-AMMP (Performance Prediction Toolkit - Analytical Memory Model with Pipelines). PPT-AMMP is a scalable, yet near-cycle accurate simulator for a serial run of a  -- typically computational physics -- code taking into account hardware features that enable instruction level parallelism, in particular pipelines and memory hierarchies. PPT-AMMP only requires actual high-level code as input and models all current, past, and even many future CPUs and cores through a large set of hardware architecture parameters scheme. Typical use cases of PPT-AMMP include (i) predicting code performance on different CPUs that a user contemplates to buy, (ii) hardware resource parameter sensitivity studies (colloquially known as ``bottleneckology''), which identify bottleneck resources such as cache sizes, cache latencies, pipeline latencies, RAM access cycles, and (iii)  hardware trend extrapolation studies that aim to predict code performance on future systems. 

PPT-AMMP at its core leverages modeling technologies that we have reported on before \cite{ppt-amm,ChennupatiSBTBM17,pypasst}. Our key contributions in this paper are (i) a \emph{fully automated nature} of our tool from high-level code $C$ as input (that does not need to be changed for our purposes) to performance prediction and (ii) a detailed \emph{pipeline model}, which is essential to accurately predict instruction-level parallelism.


The capability to directly process high-level input code $C$ coupled with scalable prediction (to any input parameter value) sets PPT-AMMP apart from all other performance prediction tools. This approach is an more user-friendly alternative to our previous modeling philosophy in PPT that requires a user to write a skeleton app mimicking the behavior of the full code (see \cite{imcsim}). Eliminating this requirement has far-reaching consequences for the PPT software stack, such as necessitating learning basic block-count functions as well as reuse distance profiles based on small-scale instances through regression; ultimately PPT-AMMP is a much more analytical-flavored model compared to the original PPT that relies more on discrete-event simulation at its core. 


We note that higher-level parallelism in threading and MPI-style communication is of course crucially important for performance prediction of HPC systems. While we believe we see a technical path towards integrating PPT's MPI and threading models \cite{pads2016,ppt} with the more scalable PPT-AMMP approach, our focus in this paper is to capture instruction-level parallelism correctly. We leave higher-level parallelism integration for future work.

The rest of the paper is organized as follows: section~\ref{sec:overview} describes the proposed system; section~\ref{sec:experiments} presents the validation experiments; sections~\ref{sec:sensitivity} shows the sensitivity analysis on cache size; section~\ref{sec:case-study} presents a case study of applying PPT-AMMP, multi-variate regression models; section~\ref{sec:background} discusses the existing literature, and finally, section~\ref{sec:conclusions} concludes and recommends future directions.

\section{Related Work}
\label{sec:background}
PPT-AMMP relies on concepts from many research areas, in particular reuse distance analysis and performance modeling.

\textbf{Reuse Distance Analysis:} Memory traces are key for an analysis on the reuse distance, however generating tera bytes worth of traces is impractical. Many attempts~\cite{Eeckhout:PAT,Zhong:PLA} tried to generate synthetic traces for approximating the reuse profiles. Although the reuse distance analysis with synthetic traces is accurate, it is nevertheless unscalable.

Other researchers used original traces~\cite{Ding:PWL,CaBetacaval:ECM}, employed sampling methods~\cite{Zhong:MRP} to identify data locality in reuse distance analysis. Probabilistic models in~\cite{Berg:SS} is another example that uses sampling for data locality prediction. Other attempts predict the cache hit-rates in parallel~\cite{Schuff:AMR} on multi-core machines.

Unlike, the earlier attempts in literature, we use smaller memory traces, sample the basic blocks, combined machine learning to extrapolate the reuse profiles of a program. Therefore, our reuse analysis offers a scalable and accurate prediction for the hit-rates and in turn the runtimes.

\textbf{Performance Modeling:} The performance modeling attempts range from analytical, simulated, empirical, and program analysis. Analytically, models such as LogP~\cite{logp,loggp} rely on human knowledge to deduce the mathematical expressions in building a model. Although these models are fast to evaluate, they pose different challenges in terms of ease of use. Simulators such as GEM5~\cite{gem5}, MARS~\cite{marss} and SST~\cite{sst} aim to produce cycle accurate predictions, for which these simulators try to execute the code on simulated target architectures. Although these simulators produce very good predictions, they are slow and unscalable.

\begin{figure}[htp]
	\centering
	\framebox{\parbox{0.95\linewidth}{
			\includegraphics[width=0.99\linewidth]{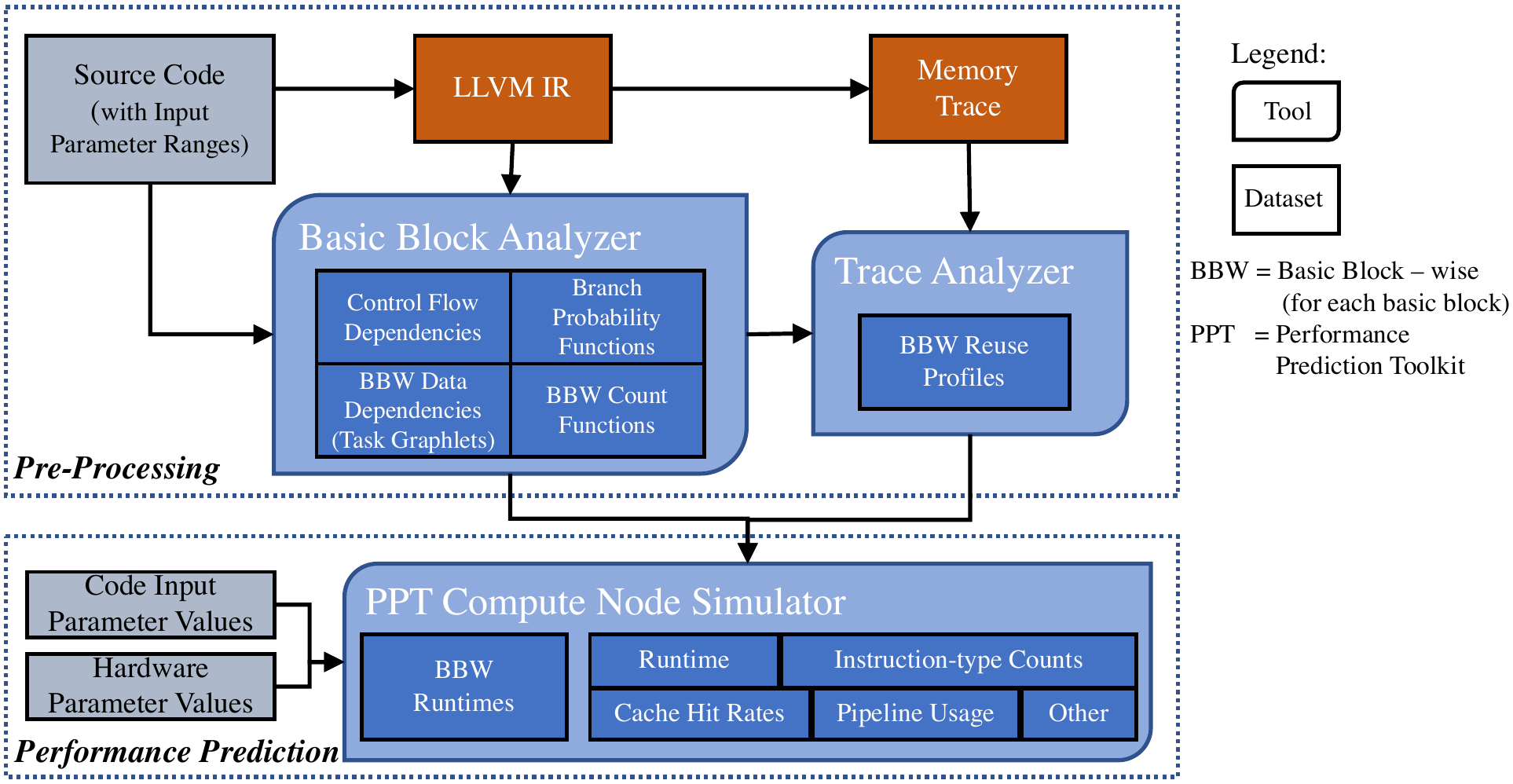}
	}}\vspace{-2ex}
	\caption{Workflow of the PPT-AMMP performance modeling}	\label{fig:ammp-design}
\end{figure}

Alternatively, program analysis approaches study the program behavior through static (compilers such as ROSE~\cite{ROSE} and Cetus~\cite{cetus}) or dynamic (for example Intel Roofline~\cite{williams2009roofline}) analysis, thereby extracting useful properties that significantly contribute towards performance. Recent attempts in program analysis include COMPASS~\cite{compass}, Durango~\cite{durango}, CODES~\cite{codes} and PyPassT~\cite{pypasst} rely on Aspen~\cite{aspen} annotations. Aspen performance modeling captures the control flow of the programs while combining the analytic modeling aspects. Note that the above methods accept source code as input and automatically produce performance models. Works in~\cite{pypasst}, applied analytical memory models and distributed performance traces to study both sequential and parallel applications. Recently, attempts in~\cite{BhimaniMLY19} used a probabilistic model with the help of regression to study the performance of parallel applications. Contrarily, attempts in Palm~\cite{palm} use annotations in a source code to produce a performance model. Palm models rely on human expertise for describing the model parameters through annotations. Other black box approaches~\cite{Lee:AER,wu2012inferred} employ regression techniques, which produce sub-optimal accuracy and pose difficult challenges to explain the behavior of the models.

Our work in this paper is in similar spirit with Aspen approaches, accepting the source code as input, then transforming it into compiler (LLVM IR) understandable representation and performs an analysis to return the final performance model and its associated predictions. However, our approach is different from the domain specific language modeling style of Aspen and its related variants, instead we employ PPT~\cite{ppt} with the Simian PDES~\cite{simian}, as the underlying rapid prototyping framework for model execution. For performance prediction, we employ PPT style task graphlets that contain basic block level data dependency graphs that are executed using simulated pipeline models.

\section{PPT-AMMP System Overview\label{sec:overview}}
Figure~\ref{fig:ammp-design} shows the overall design of AMMP. At the highest level, AMMP consists of a pre-processing stage (top part of the figure) and a performance prediction stage (lower part).

The pre-processing stage is executed once for a given source code $C$, which for example is {\tt MatrixMultiply}. PPT-AMMP can take any high-level language (and for now serial) code as input as long as it can be translated into LLVM, a standard compiler infrastructure that supports C, C++, and Fortran. The pre-processing stage analyzes the code, stores the result in data sets, which become inputs to the {\em Compute Node Simulator} in the performance prediction stage. The main tools in the pre-processing stage are {\em Basic Block Analyzer} and {\em Trace Analyzer}, detailed in sections~\ref{sec:bbanalyzer} and~\ref{sec:traceanalyzer}. The pre-processing stage is strictly \emph{architecture-independent}, a key-characteristic; this enables us to predict performance in the second stage for an arbitrary hardware platform. In addition to the code $C$, we need a list of input parameters $|I|$ to $C$ that may affect performance, for which we create a set of small-instance runs of $C$. 

The performance prediction stage requires as input the size $|I|$ of the input $I$ to code $C$; thus with $C =$ {\tt MatrixMultiply}, we would have $I$ to be the two input matrices of size, say $n \times l$ and $l \times m$, thus $|I| = \{m, l,n \}$, i.e. the three matrix size parameters. Next, we specify the hardware parameter values $H$ (such as clock speed, cache sizes, and latencies, instruction pipelines and latencies, memory size, and more);  this is done through an input file that contains all values for e.g., an Intel Haswell processor (PPT contains a library of most modern processors). The main tool in the performance prediction stage is the PPT Compute Node Simulator, accepts the pre-computed code analysis data sets from the Basic Block Analyzer and the Trace Analyzer and the input values $|I|$ and $H$, with which calculates the runtime, as well as many other performance features, such as cache hit rates, pipeline usage, or instruction-type counts. We describe the PPT Compute Node Simulator in section~\ref{sec:pptcomputenode}. 

\subsection{Basic Block Analyzer}
\label{sec:bbanalyzer}
The {\em Basic Block Analyzer} (BBA) analyzes the structure of the input code $C$ at an individual basic block level. A \emph{basic block} is a straight-line code with a single entry and a single exit, with no intermediate branches except for a possible branch at the exit; the concept of a basic block is standard in compiler design. The Basic Block Analyzer works in the following steps.

\begin{figure}[htp]
	\centering
	\begin{minipage}{0.47\textwidth}
		\footnotesize
		\centering
		\begin{lstlisting}[language=C]
			#define N 256;
			float** r8_ijk(float** a, float** b, float** c) { 
				int i, j, k; //Initialization
				for (i = 0; i < N; i++) //i loop
				for (j = 0; j < N; j++) { //j loop
					a[i][j] = 0.0; 
					b[i][j] = c[i][j] = 1.0; 
				}
				for (i=0; i<N; i++) //i loop
				for (j=0; j<N; j++) //j loop
				for (k=0; k<N; k++) //k loop
				a[i][k] = a[i][k] + b[i][j] * c[j][k];
				return a;
			} //end of r8_ijk()
			int main() {
				float A[N][N], B[N][N], C[N][N];
				A = r8_ijk(A,B,C);
				return 0;
			} //end of main()
		\end{lstlisting}
		\vspace{-2ex}\caption{Matrix multiplication as an input code $C$ example, only relevant function shown}\label{fig:ex}
		\vspace{-1em}
	\end{minipage}\hfill
	\begin{minipage}{0.50\textwidth}
		\centering
		\framebox{\parbox{0.70\linewidth}{
				\includegraphics[width=0.99\linewidth,height=0.99\linewidth]{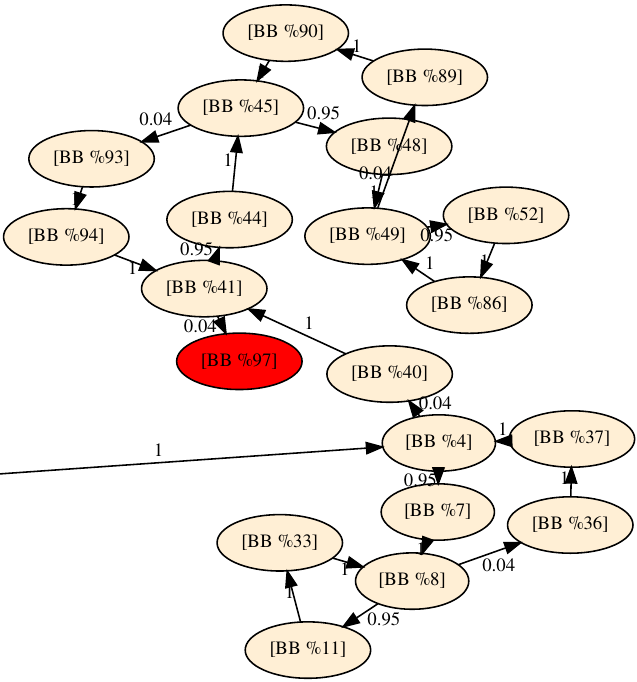}
		}}
		\vspace{-1ex}\caption{Control Flow Graph $CFG^C$ of the basic blocks in {\em r8\_ijk}. The numbers on the vertices indicate execution counts, the edge labels are branching probabilities} \label{fig:cfg}
	\end{minipage}
\end{figure}

\subsubsection{Compute Control Flow Dependencies} 
Input code $C$ is first compiled into an architecture-independent LLVM Intermediate Representation (IR), using a compiler, such as \texttt{clang} for C or \texttt{flang} for Fortran. From the IR, BBA identifies the set of, say $m$  basic blocks $BB^C = \{ BB_1, \ldots, BB_{m}\}$ and builds the control flow graph  $CFG^C =  (BB^C, E^C)$ with vertex set $BB^C$ and directed edge set $E^C \subset BB^C \times BB^C $. Figure \ref{fig:ex} shows an example of $C$ for {\tt MatrixMultiply}.

Figure~\ref{fig:cfg} shows the $CFG^C$. The original source code $C$ is sequential in nature with one function ({\em r8\_ijk}) call from the {\em main} function. The function, {\em r8\_ijk} contains the initialization of data elements in a nested {\em for} loop, while the actual naive matrix multiplication is in a triple nested {\em for} loop. In this example, the CFG of matrix multiplication contains a total of $22$ basic blocks. For ease of illustration, we omitted the basic blocks in the {\em main} function, all the other basic blocks belong the {\em r8\_ijk} function. In general, a {\em for} loop is decomposed into $5$ basic blocks: initialization, condition, body, increment and {\em for} end. The directed cycles in the graph represent these {\em for} loops in the original source code. With some compiler optimizations, the number of basic blocks can be reduced significantly. The red colored node is the terminating block of the function.

\subsubsection{Compute Branch Probability and Basic Block Count Functions} 
\label{sec:bb_count_pred}
The core feature of our approach is our analysis along basic blocks. In this step, we compute branching probabilities $p_{i,j}$ for each edge $(BB_i, BB_j) \in E^C$ of the control flow graph and execution counts $N_i$ for each basic block $BB_i \in BB^C$. Figure \ref{fig:cfg} shows an example of these branching probabilities and execution counts for a fixed input size (in this case two 4x4 matrices). BBA calculates these probabilities and counts as {\em functions} of the input parameters $I$.

To this end, BBA runs a fully factorial experimental design on a small number of performance-relevant input values (which need to be identified by the user). To be more precise, we let the set of input parameters be $I = \{i_1, \ldots, i_{|I|}\}$ (e.g., the three matrix dimensions in matrix multiplication); we define a small set of test values for each $i_j$; say $k<5$ different values for the matrix dimensions in our example with small values, e.g. $\{1, 2, 4, 8\}$. We run all $k^{|I|}$ input parameter combinations of code $C$ with automated LLVM-level instrumentation to get the desired branching probabilities and execution counts for these small instances. 
BBA then uses a multi-linear regression as a machine learning approach to fit these functions. Multi-linear regression represents the predictor variable in the form of one or more independent variables. For example, in matrix multiplication, matrix sizes are our dependent variables while the number of times that each of the total basic blocks execute is our predictor. For a program with {\em N} basic blocks, we will prepare {\em N} regression models each of which is a combination of input variables. Therefore, Eq.~\ref{eq:reg} is the general form of regression.
\begin{equation}\label{eq:reg}
	N_i = \alpha_1(i_1\times i_2\dots i_k) + \alpha_2(i_2\times i_3\dots i_k) + \dots + \alpha_k i_k + c 
\end{equation}
\noindent where $\alpha_1$, $\dots$, $\alpha_k$ are weights; $i_1$, $\dots$, $i_k$ are the program specific inputs (independent variables); $c$ is a constant and $N_i$ (the dependent variable) is the number of executions of $i^{th}$ basic block. In order to minimize any overfitting effects, we regularize the model in Eq.~\ref{eq:reg} with 1-norm on the input variables. A similar approach is used for the branching probabilities $p_{i,j}$.

\subsubsection{Compute Data Dependency Graphlets for each Basic Block.} 
For each basic black $BB_i$, the BBA tool builds a data dependency graph $DDG_i^C  = (O_i, E_i)$, where $O_i$ is the set of operation instructions in $BB_i$ and a directed edge $(v,w)$ of two instructions $v,w \in O_i$ is in $E_i$ if the child instruction $w$ accesses a data element that the parent instruction $v$ accesses as well. We speak of graphlets (as opposed to graphs) because basic block typically contains a handful of instructions. The key insight for the $DDG_i^C$ graph is that vertices in $O_i$ are executed as soon as the parent vertices have been executed independent of their original order in the code $C$. Modern compilers as well as hardware architecture aim to exploit such instruction level parallelism opportunities to fill their instruction pipelines. 

More precisely, BBA parses through each basic block in the LLVM IR to generate the task-graphlets. The traversal is in the order of the control flow (see Figure~\ref{fig:cfg}) of execution of basic blocks. 

\begin{figure}[htp]
	\centering
	\begin{minipage}{0.45\textwidth}
		\begin{lstlisting}[language=llvm]
			; <label>:6            ;preds = %3
			%7 = load i32, i32* %i, align 4
			%8 = mul nsw i32 2, %7
			%9 = load i32, i32* %i, align 4
			%10 = sext i32 %9 to i64
			%11 = load i32*, i32** %1, align 8
			%12 = getelementptr inbounds i32, i32* %11, i64 %10
			store i32 %8, i32* %12, align 4
			br label %13
		\end{lstlisting}
		\vspace*{-1ex}\caption{Annotated BasicBlock from LLVM IR}\label{fig:llvm-ir}
	\end{minipage}\hfill
	\begin{minipage}{0.52\textwidth}
		\framebox{\parbox{0.90\linewidth}{
				\begin{tikzpicture}[>=stealth,shorten >=3pt,auto,node distance=1cm]
					\node[shape=circle,draw=black] (A) at (1,1.6) {8};
					\node[shape=circle,draw=black] (B) at (0.4,1.1) {4};
					\node[shape=circle,draw=black] (C) at (3.3,-0.2) {2};
					\node[shape=circle,draw=black] (D) at (0.5,0) {9};
					\node[shape=circle,draw=black] (E) at (3,1.6) {3};
					\node[shape=circle,draw=black] (F) at (-0.8,1.3) {5};
					\node[shape=circle,draw=black] (G) at (-2.0,-0.4) {6};
					\node[shape=circle,draw=black] (H) at (-2.4,1.85) {7};
					\node[shape=circle,draw=black] (I) at (4.2,1.2) {1};
					
					\path [->, thick] (A) edge node {} (D);
					\path [->, thick] (C) edge node {} (D);
					\path [->, thick] (E) edge node {} (D);
					\path [->, thick] (B) edge node {} (D);
					\path [->, thick] (F) edge node {} (D);
					\path [->, thick] (H) edge node {} (D);
					\path [->, thick] (G) edge node {} (D);
					\path [->, thick] (C) edge node {} (E);
					\path [->, thick] (E) edge node {} (A);
					\path [->, thick] (F) edge node {} (H);
					\path [->, thick] (G) edge node {} (H);
					\path [->, thick] (B) edge node {} (F);
					\path [->, thick] (H) edge node {} (A);
					\path [->, thick] (I) edge node {} (C);
				\end{tikzpicture}
		}}
		\vspace{-1ex}\caption{Intra-dependency graph of instructions in a BB.} \label{fig:graphlet}
	\end{minipage}
\end{figure}

Figure~\ref{fig:llvm-ir} shows the example basic block $BB_6$ extracted from the LLVM IR of matrix multiplication. Figure~\ref{fig:graphlet} shows the corresponding {\em task-graphlet} $DDG_6$. Each node in the {\em graphlet} represents an instruction while the edges stand for the dependencies among the instructions in that basic block. In this example, the nodes with inbound arrows represent that this particular instruction will be executed after completion of all the instructions in the inbound nodes. 

For example, \texttt{node 9} is a {\em branch} (\texttt{br}) instruction which will only be taken upon the execution of all the other preceding instructions in the basic block. In Figure~\ref{fig:graphlet}, the numbers in the nodes represent the corresponding line numbers (\texttt{1} -- \texttt{9}) of the instructions (refer to Figure~\ref{fig:llvm-ir}) in the basic block. In the intra-dependency graph: \texttt{node 1} stands for basic block entry while \texttt{node 9} is the \texttt{br} statement and executed upon exit; \texttt{nodes 2}, \texttt{4}, \texttt{6} stand for three {\em load} operations, which are executed independently through the dedicated pipeline port for {\em load} and {\em store}; The \texttt{node 3} is an integer multiply ({\em mul}) of a variable (dependent on the \texttt{node 2}) with a constant -- which has a dedicated port in the simulated pipeline model and will only be executed upon resolving the dependencies; The \texttt{nodes 5} and \texttt{7} represent type casting and getting the address of an element, which do not really have any compute significance, therefore, we omit this type of instructions; Although, \texttt{node 8} is a {\em store} instruction and has a dedicated port in the pipeline model, the dependencies on \texttt{nodes 3} and \texttt{7} make it wait from executing independently.

Summing up, the Basic Block Analyzer BBA - upon input of a source code $C$ -- thus prepares the following data items for use by the Compute Node Simulator and the Trace Analyzer:
\begin{enumerate}
	\item Control Flow Graph $CFG^C$
	\item Basic Block Count Functions $N_i$ for each basic block $BB_i \in BB^C$ as a function of input parameters $I$
	\item Branching Probability Functions $p_{i,j}$ for each edge $(BB_i, BB_j) \in E^C$ of $CFG^C$
	\item Data Dependency Graphlets $DDG_i^C  = (O_i, E_i)$ for each basic block $BB_i$
\end{enumerate}

\subsection{Trace Analyzer}
\label{sec:traceanalyzer}
The memory hierarchy with different cache levels as well as main memory is a key factor in determining runtime performance of an input code $C$. Its dynamics is extraordinarily difficult to model correctly without resorting to explicit simulation of memory load and store into an execution trace. The key architecture-independent concept of {\em Reuse Distance} is a reasonably well-studied albeit compute intensive measure that characterizes the structure of memory accesses of an input code $C$. Reuse distance is the number of unique memory references between two consecutive accesses of the same reference. If we look at the reuse distances of all memory addresses touched in an execution trace of a code $C$ on input parameters $I$, all these individual reuse distances form a histogram, which we call {\em reuse profile}. Through well-known formulas (explained in section~\ref{sec:pptcomputenode}), we get the architecture-dependent cache hit-rates from the reuse profile. These cache hit-rates in turn determine the effective duration of the memory access instructions in $C$.

Our Trace Analyzer tool calculates the reuse profiles on an individual basic-block as a function of code input parameters $I$. The reuse profile $P_i(d, I)$ for basic block $BB_i$ and reuse distance $d$ is the relative frequency (i.e. actual counts normalized by the total number of memory accesses) of encountering memory accesses with reuse distance $d$ in code $C$ with input parameters $I$. In practice, distances $d$ can be binned into a few categories. Our previous work~\cite{ChennupatiSBTBM17} showed that the per-basic-block approach to reuse distance works. Our focus in this paper is to show how to integrate it into the larger PPT-AMMP system through our Trace Analyzer.


In order to generate the reuse profiles, we  produce a memory trace (set of memory references generated in the order of sequence of execution) of a the input code $C$. While other approaches for reuse profile in literature~\cite{Ding:PWL,Fang:RMP} use large memory traces for reuse distance analysis, we produce small traces in order to guarantee scalability in predictions. Similar to the basic block execution count estimates through multi-scale regression, for a small $k<5$ and small input parameter values. We learn the reuse profile as functions of $I$ through multi-linear regression.

We give a few more implementation details. 
In order to record the memory references dynamically, we instrument the source code. We implemented an LLVM-based source instrumentation tool, which associates the run-time memory references to the correct basic blocks. The annotated memory trace generated by this instrumentation is passed through a post-process step to handle possible function calls which appear in various basic blocks, and then those basic blocks in a recursive fashion. Our traces are architecture independent, similar to attempts in~\cite{Steen:UGhent} which show architecture independent performance in energy modeling.
\vspace{-2ex}
\begin{algorithm}[htp]
	\footnotesize
	\begin{algorithmic}[1]
		\Procedure{$reuse\_profile\_BB_i$}{$BB_i$, $memory\_trace$}
		\State $reuse\_distances$, $sampled\_wins \gets [\;]$, $[\;]$
		\State $sample\_size \gets x$ \Comment{$x\%$ of all the $BB_i$(s)}
		\For{$bb$ {\bf in} $all\_BB_i$}
		\State $sampled\_wins.append([BB_i\_start,BB_i\_end])$ 
		\EndFor
		\State $windows \gets random(sampled\_wins, sample\_size)$ \Comment{Random sampling with a {\em sample\_size}}
		\For{{\em window} {\bf in} {\em windows}}
		\For{{\em addr} {\bf in} {\em memory\_trace[window]}}
		\State $reuse\_dist$ $\gets$ {\em max\_back\_reference}({\em addr}) \Comment{Look for this address back in the trace and the unique addresses between the two memory accesses}
		\EndFor
		\State $reuse\_distances.append(reuse\_dist)$
		\EndFor
		\State $uniq\_reuse\_dist, counts \gets$ unique($reuse\_distances$)
		\State $prob\_rd \gets {\bf map}({\bf lambda}$ $x$: $x/len(reuse\_distances)$, $counts)$ 
		\State $r\_prof_i \gets$ {\bf zip}$(uniq\_reuse\_dist$, $prob\_rd)$ 
		\State \Return $r\_prof_i$
		\EndProcedure
	\end{algorithmic}
	\caption{Reuse profile $P_i$($d,I$) of basic block $BB_i$, given trace produced by input parameters $I$ }\label{alg:sdprof}
\end{algorithm}\vspace{-2ex}

Algorithm~\ref{alg:sdprof} presents the calculation of conditional reuse profile of a basic block.
Given the trace of a basic block, we measure the reuse distance for each memory reference. Starting from the current memory address, we look back in the trace (across the basic blocks) for the same address (termed {\em max\_back\_reference}). The number of unique other references between the above two accesses is the reuse distance of that reference in the basic block. In case, if the {\em max\_back\_reference} is absent from the trace then the reuse distance becomes infinite ($\infty$) We repeat this for each address in the basic block, recording all the reuse distances. For each of these reuses, we count the number of occurrences from which the corresponding conditional reuse distance probabilities are calculated. The pair of these conditional reuse distances and the corresponding probabilities for the conditional reuse profile are recorded. 

\subsection{Performance Prediction Toolkit (PPT) Compute Node Simulator}
\label{sec:pptcomputenode}
\subsubsection{Overview}
The performance prediction stage of PPT-AMMP is executed in the PPT Compute Node Simulator, referred as simulator, here after for brevity. The simulator traces its lineage to the Performance Prediction Toolkit (PPT)~\cite{ppt,ppt-amm,pads2016,tompecs2016} and in fact represents a special use case as we focus on serial input code, whereas PPT has its traditional strengths in MPI modeling. 

The simulator takes the data produced from the Basic Block Analyzer and the Trace Analyzer for code $C$ as input; recall that these data sets are (i) the control flow graph, (ii) the count functions for each basic block, (iii) the branching probability function for each edge of the control flow graph, (iv) the data dependency graphlets for each basic block, and (v) the reuse profiles for each basic block. In addition, it takes as input a set of $C$-input size parameters $|I|$ (e.g., the matrix dimensions for matrix multiplication) and a set of hardware architecture parameters $H$. Of course, hardware architecture parameters can be pre-stored, in fact the open sourced PPT~\cite{ppt} contains a library of most existing processor models. A list of hardware architecture parameters is shown in Table ~\ref{tab:hardware}; a user can specify these parameters.

\begin{table}[htp]
	\centering
	\caption{Hardware Architecture Parameters $H$} \vspace{-1ex}
	\begin{tabularx}{\columnwidth}{|l|l|X|}
		\hline
		\textbf{ Category} &\textbf{ Parameter } & \textbf{Explanation}\\ \hline
		General & Clock speed & in Hz \\ \hline
		Pipeline & Instruction Set & Group into instruction types, e.g.,iALU, fALU, fDIV, mov, etc.\\ \hline
		& Pipeline Counts & Number of pipelines per instruction type \\ \hline
		& Latency& for each instruction pipeline\\ \hline
		& Throughput& for each instruction pipeline\\ \hline
		Memory & Cache Level Count& Number of cache levels  \\ \hline
		& Size&  List of sizes for each cache level\\ \hline
		& Latency& for each cache level \\ \hline
		& Bandwidth& for each cache level \\ \hline
		& Associativity&  for each cache level \\ \hline
		& Line Size&  for each cache level \\ \hline
		& RAM Size&   \\ \hline
		& RAM Latency&   \\ \hline
		& RAM Bandwidth&   \\ \hline
	\end{tabularx}
	\label{tab:hardware}
	\vspace{-0.5em}
\end{table}

Referring back to Figure~\ref{fig:ammp-design}, the Compute Node Simulator outputs the predicted runtime $t_i$ for each basic block $BB_i$. This is the main computational step of the simulator. The detailed discrete event simulation goes through each instruction of the dependency graphlet $DDG_i^C$ in a step-wise fashion mimicking pipeline behavior and memory hierarchy level misses (based on the reuse profiles $P_i(d, I)$. Note, however, that in total we only need to execute each instruction once (independent of how often the real code would execute it), thus making our approach scalable. 

\begin{figure*}[htp]
	\centering
	\includegraphics[width=0.95\linewidth,height=0.35\linewidth]{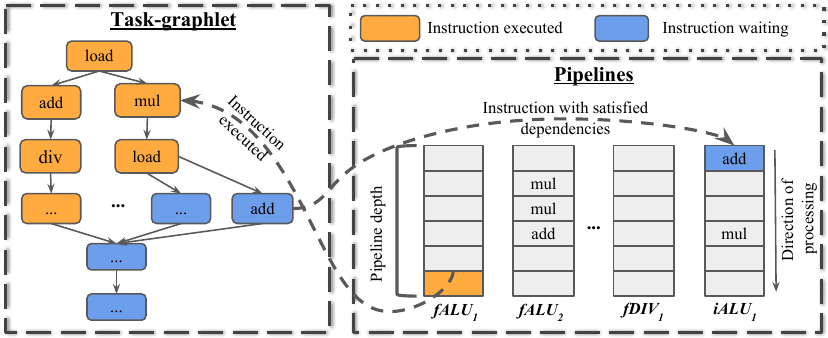}
	\vspace{-1ex}\caption{General pipeline model for a target architecture} \label{fig:pipe-hw}
\end{figure*}
Next, we calculate the predicted overall run-time $T_H^C(|I|)$ of code $C$ with input parameters $|I|$ on hardware platform $H$: 
\begin{equation}
	T_H^C(|I|) = \sum_{i=1}^m t_i \times N_i(|I|) 
\end{equation}
\noindent where the sum is over all $m$ basic blocks and $N_i(|I|)$ is the execution count of basic block $BB_i$ given input parameters $|I|$ (we dropped the indices $C$ and $H$ for the constituents of the formula for simplicity). Similarly, we can compute the overall distribution of instruction type counts, pipeline usage statistics, and similar metrics. 

We now describe the prediction of basic block level runtimes $t_i$, which involves two steps: (i) calculating the effective memory latency (we calculate for the entire computation rather each individual basic block) and (ii) simulating the instruction pipelines.

\subsubsection{Calculating Effective Memory Latency}
In order to calculate an effective average response time for a memory operation, we first estimate the {\em reuse profile} of the entire code $C$ on input $|I|$ that we call $P^C(d,|I|)$, as shown in Eq.~\ref{eq:sdprog}
\begin{equation}\label{eq:sdprog}
	P^C(d,|I|) = \frac{1}{N(|I|)}\sum\limits_{i=0}^{m} N_i(|I|) \times P_i(d,|I|)
\end{equation}
\noindent where, $d$ is the reuse distance, $m$ is the number of basic blocks, $N_i(|I|)$ is the execution count of block $BB_i$, and $P_i(d,|I|)$ is the reuse profile for block $BB_i$ (note, both are measured in the pre-processing step). The term $N(|I|) = \sum_{i=1}^m N_i(|I|)$, is the sum of all basic block counts. Also, $P^C(d,|I|)$ is still architecture independent.

In order to get to the average memory instruction latency, we estimate cache hit-rates for each level. We will work our way backwards: assuming we have cache hit-rates, we predict effective latency and (reciprocal) throughput or bandwidth using Eq.~\ref{eq:avg_latency}.
\begin{equation} \label{eq:avg_latency}
	\begin{aligned}
		\lambda_{eff} = P_{L_1}(h) \times \lambda_{L_1} + \bigl(1-P_{L_1}(h)\bigr)\biggl[P_{L_2}(h) \times \lambda_{L_2} + \bigl(1-P_{L_2}(h)\bigr)\\
		\bigl[P_{L_3}(h) \times \lambda_{L_3} + \bigl(1-P_{L_3}(h)\bigr)\times \lambda_{RAM}\bigr]\biggr]
	\end{aligned}
\end{equation}
\noindent where, $\lambda_{L_*}$ and $\lambda_{RAM}$ are the hardware latencies of different caches and $RAM$ respectively;  $P_{L_*}(h)$ represents the hit-rate at different caches, calculated using Eq.~\ref{eq:phit} below, where $h$ is the event of a cache hit. Similarly, we calculate the average effective bandwidth, $\beta_{eff}$ (replace $\lambda$s in Eq.~\ref{eq:avg_latency} with $\beta$).
\begin{equation}\label{eq:phit}
	P(h) = \sum\limits_{i=0}^{i_{max}} P^C(d_i,|I|) \times P({h\mid d_i})
\end{equation}
\noindent where, $P^C(d_i,|I|)$ is the probability of $i^{th}$ reuse distance bin (e.g., non-zero value) in the reuse distribution $P^C(d_i,|I|)$, and $P({h\mid d_i})$ is the probability of a hit for a memory instruction with a reuse distance of $d_i$. 

The mixture model for finding the hit-rates $P({h\mid d_i})$ at a given reuse distance is shown in Eq.~\ref{eq:phd}, derived from the {\em stack distance based cache model} (SDCM)~\cite{brehob:analytical} to estimate cache hit-rates.
\begin{equation}\label{eq:phd}
	P(h\mid d) =  \sum\limits_{a=0}^{A-1}\binom{d}{a}\biggl(\dfrac{A}{B}\biggr)^a\biggl(\dfrac{B-A}{B}\biggr)^{(d-a)}
\end{equation}
\mbox{where} {\em d} is the reuse distance, {\em A} is the associativity and {\em B} is cache size in terms of number of blocks (cache size over cache line size). While actual hardware implementations of caching hierarchies may differ a bit (and are often proprietary), Eq.~\ref{eq:phd} is generally accepted as a good approximation for hit rates and widely used.

Thus, the Compute Node Simulator calculates the effective memory instruction latency $\lambda_{eff}$ and bandwidth $\beta_{eff}$  as outlined through the equations above; the computational complexity is $O(m i_{max}$, where $i_{max}$ is the number of reuse distance bins (usually less than 20). Note that both values depend on the architecture parameters $H$. The resulting $\lambda_{eff}$ and  $\beta_{eff}$ become parameters of the pipeline for memory operations that we describe in the next section.

\subsubsection{Simulating Instruction Pipelines}
The PPT Compute Node Simulator uses the Simian~\cite{simian} Parallel Discrete Simulation Engine, written in Python. Application processes or threads are implemented as co-routines. If co-routines yield to other co-routines this results in an event, often called as process-based simulation.

The simulator calculates the basic block execution times $t_i$ for each basic block using the data dependency graphlets $DDG^C_i$.
It processes each data dependency graphlet   $DDG_i^C = (O_i,E_i)$ in the following manner: 
The main simulator process checks for each vertex $v \in O_i$, whether its parent vertices have already completed their pass through the pipeline. It then enqueues each such instruction vertex into the most immediately available appropriate pipeline. A hardware pipeline is an architectural construct which is intended to reuse circuit elements for multiple instructions, thus achieving a form of parallel efficiency. Modern processors can contain several pipelines for each type of instruction. Each pipeline is characterized by a pipeline-latency parameter $\lambda_P$, which is the number of clock cycles required to move a single instruction through the pipeline, and a pipeline throughput parameter $\beta_P$, which is the number of clock cycles the pipeline takes to move an instruction to its next pipelining stage. A pipeline can be viewed as a multi-stage waiting queue consisting of say $k$ different stages (sometimes called the pipeline depth since $\lambda_P = k \beta_P$). The logic thus is as follows: if only a single add instruction is inserted into a pipeline, it will take $\lambda_P$ clock cycles to be completely processed (we call this ``latency-bound''); if, say 2 add instructions are inserted consecutively into the same pipeline, the second instruction will be completed after $\lambda_P+\beta_P$ clock cycles; if $k$ add instructions are inserted consecutively, the total time will take $\lambda_P+k\beta_P$ (full pipeline). A data dependency graphlet that uses the available pipeline structure well, will result in a nearly full pipeline during its execution. Note that the throughput and latency parameters from memory operations computed through the re-use distance analysis flow into a special memory operations pipeline (that acts like any other pipeline).

Figure~\ref{fig:pipe-hw} shows a conceptual overview of how the a graphlet gets processed and put into the pipelines.
The instruction vertices in the data dependency graphlet are processed in topological order, based on data-dependency.
The main loop of the simulator is the following for each basic block graphlet $DDG_i^C = (O_i,E_i)$: 
\begin{enumerate}
	\item For each instruction vertex $v \in O_i$, check whether all its parents vertices $u$ with $(u,v) \in E_i$ have already been marked as executed. Call such v
	\item 
	For each vertex $v$ (whose parent dependencies have been executed), find the next available pipeline that can handle the instruction-type of $v$ and add it that local queue.
	\item 
	If a pipeline finishes execution of an instruction vertex $v$, mark the vertex as ``executed'' and return to Step 1 (making other instruction vertices eligible for execution).
	\item
	Report final time $t_i$
\end{enumerate}
Each pipeline runs as a separate simulator process that accepts instructions $v$ at a simulated time and then moves this instruction through its pipeline stages at the throughput speed $\beta_P$ per pipeline stage. Once $v$ has reached the end of the pipeline, the pipeline process informs the main simulator process that it has completed.




\section{Validation Experiments}
\label{sec:experiments}
We validate the proposed system, AMMP, on four benchmark problems: {\em JACOBI}, {\em MATMUL}, {\em LAPLACE2D} and {\em BlackScholes}. 

We validate and illustrate PPT-AMMP along the major components of the Basic Block Analyzer, Trace Analyzer, and the Compute Node Simulator with their associated outputs.

The target architecture in our experiments is an \texttt{Intel Xeon E5-2695} processor with a clock speed of $2.1$ GHz. The target architecture has a three level cache hierarchy cache with sizes $L_1$ = $64$K, $L_2$ = $256$K, $L_3$ = $46080$K, with $L_3$ being shared across all the cores. The associativity of all three hierarchies are $8$, $8$, $20$ respectively. The compiler used in the experiments is \texttt{clang v3.9.0}, both with (\texttt{-O3}) and without optimization (\texttt{-O0}) flags.

\subsection{Experimental Setup}
In the pre-processing step, we use standard versions of our benchmarks to PPT-AMMP, together with clear indication of what input parameters $I$ should be varied, which in these simple cases is always just a single parameter, namely matrix size or data set size.

\begin{figure}[htp]
	\footnotesize
	\centering
	\begin{lstlisting}[language=Python]
		pipeline_latencies = { # in seconds
			'iadd' : 1.04e-9, 'fadd': 1.3e-9, 'idiv': 9.46e-9, 'fdiv': 15.46e-9, 'imul': 1.54e-9, 'fmul': 2.31e-9,'load': 0.38e-9, 'store': 0.38e-9, 'alu': 0.38e-9, 'br': 0.38e-9, 'unknown': 0.38e-9}
		
		pipeline_throughputs = { # in seconds
			'iadd' : 0.25e-9, 'fadd': 0.38e-9, 'idiv': 3.46e-9, 'fdiv': 3.07e-9, 'imul': 0.5e-9, 'fmul': 0.36e-9,'load': 0.38e-9, 'store': 0.38e-9, 'alu': 0.38e-9, 'br': 0.38e-9, 'unknown': 0.38e-9}
	\end{lstlisting}
	\vspace{-2ex}\caption{Hardware parameters for pipeline instructions}\label{fig:pipe-params}\vspace{-1em}
\end{figure}

For the second stage, performance prediction, we populate the hardware parameters as outlined in Table \ref{tab:hardware}. 
Figure~\ref{fig:pipe-params} shows the hardware parameters (latency and throughput) for various instructions in a pipeline. Along with these parameters, we use the estimated (through reuse profiles) effective latency and throughput required for a memory access in predicting the final runtime of an application. We use data from the  Agner Fog manual~\cite{Agner:Inst} to populate the pipeline hardware parameters.

\subsection{The Basic Block Analyzer}
In this section, we present the functioning of the Basic Block Analyzer (BBA). Recall that the BBA calculates the control flow graph of the basic blocks $CFG$, the data dependency graphs within each basic block $DDG_i$, the branch probabilities and the basic block execution count functions $N_i$ are modeled as functions of input size. For example, $N_i$ = $f(X)$, where $X$ is a vector of input parameters to an application. In fact, the BBA uses a static analysis on the LLVM IR of the source codes in order to extract the DDG, basic block counts, etc. These counts and graphs are essential in order to deduce the extrapolating fits for each of the basic block as well as to exercises the pipeline model.

\begin{table}[htp]
	\centering
	\caption{Number of basic blocks $BB_i$ present in the program and the kernel of the program}\label{tab:graphlets}\vspace{-2ex}
	\begin{tabular}{|c|c|c|c|c|c|}
		\toprule
		\multirow{2}{*}{\textbf{Benchmark}} & \multicolumn{2}{c|}{\textbf{\# program blocks}}  & \multicolumn{2}{c|}{\textbf{\# kernel blocks}} \\
		\cline{2-5}
		& \textbf{Unopt} & \textbf{Opt} & \textbf{Unopt} & \textbf{Opt}  \\
		\midrule
		{\em JACOBI} & 97 & 70 & 28 & 9 \\
		{\em MATMUL} & 22 & 12 & 8 & 4 \\
		{\em LAPLACE2D} & 26 & 9 & 6 & 1 \\
		{\em BlackScholes} & 54 & 44 & 17 & 8 \\
		\bottomrule
	\end{tabular} 
\end{table}
\begin{figure}[htp]\vspace{-2ex}
	\centering
	\subfloat[Unoptimized]{{\includegraphics[width=0.47\linewidth, height=0.38\linewidth]{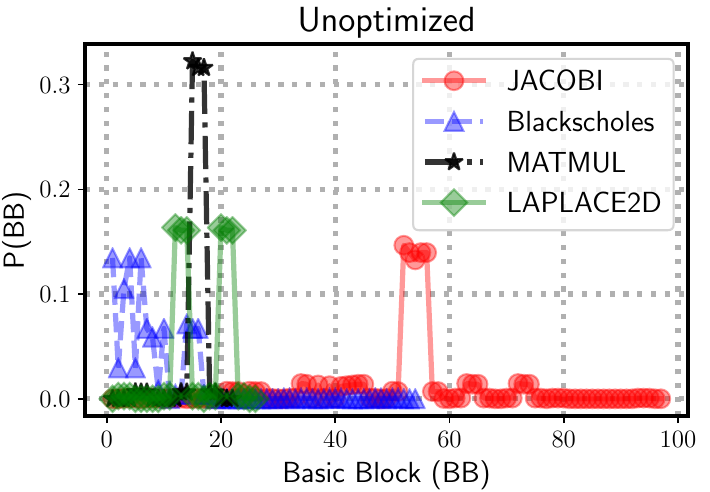} }}%
	\qquad
	\subfloat[Optimized]{{\includegraphics[width=0.47\linewidth, height=0.38\linewidth]{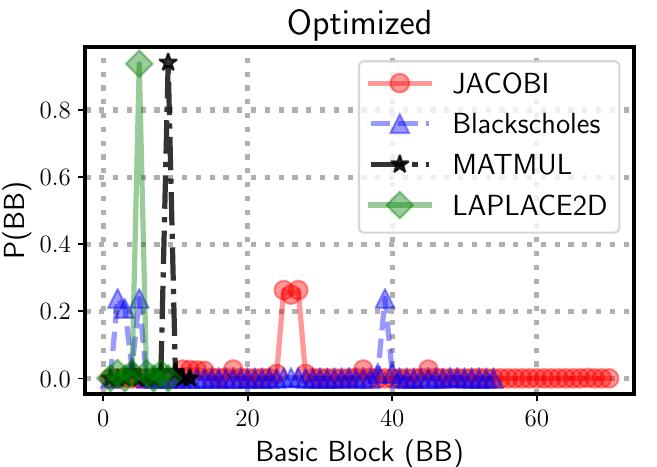} }}
	\vspace{-2ex}
	\caption{Discrete probability distributions of basic block executions of all the four benchmarks (a) without and (b) with optimizations}
	\label{fig:bb_probs} \vspace{-2ex}
\end{figure}
\begin{figure}[htp]
	\centering
	\includegraphics[width=.85\linewidth]{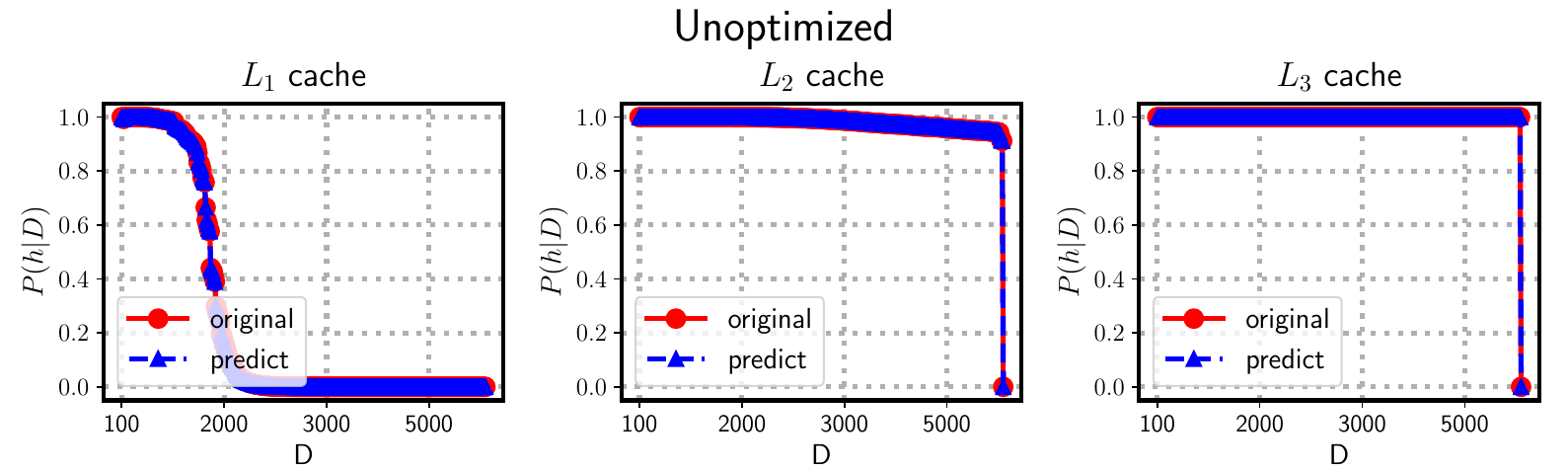}
	\includegraphics[width=.85\linewidth]{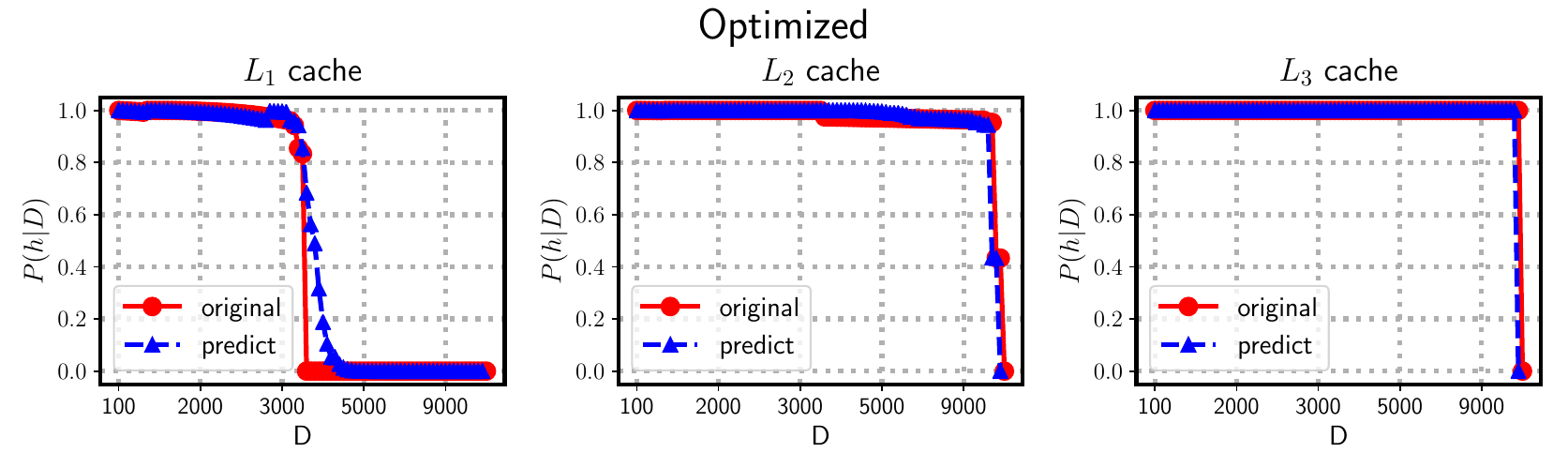}
	\vspace{-3ex}
	\caption{Cache hit-rates at a given reuse distance for both unoptimized and optimized versions of {\em JACOBI}}
	\label{fig:sd_phit}\vspace{-2ex}
\end{figure}
The basic block structure is generated from the LLVM IR code. Table~\ref{tab:graphlets} presents a detailed analysis on the number of basic blocks with and without optimizations for all the benchmarks. We report two different statistics on the number of basic blocks -- entire program and the kernel of the program. Note the kernel in this context represents the set of basic blocks that have most significant contribution towards the program execution. Of the four benchmarks, {\em JACOBI} has the highest number of basic blocks with and without optimizations. The count of the basic blocks (graphlets) depends on the original source code despite the optimizations. 

Figure~\ref{fig:bb_probs} shows the probability distributions of executing all the basic blocks of a given program for all the four benchmarks; these are derived from the $N_i$ values from our earlier description. The distributions are discrete, while the peaks in the distribution represent the basic blocks that are executed more often, therefore having significantly higher influence on the runtimes. Note, these peaks belong to the kernel of a benchmark, which are highly correlated (in both the cases of with and without optimizations) to the number of kernel blocks. For example ({\em JACOBI} and {\em BlackScholes}), a few number of basic blocks have approximately zero probability due to the fact that they are executed only once and/or never executed. The contribution of such basic blocks towards runtime is negligible. We used these discrete distributions in extrapolating the conditional reuse profiles of the most significant basic blocks. 

To see the result of the multi-regression technique to find the growth patterns of the block execution counts $N_i$, please refer to Table \ref{tab:bbtimes}, which shows for instance the function of $N_i$ for  {\em JACOBI} to be \begin{math}N_4^{JACOBI}(n) = n^2-3n+2\end{math}.

\subsection{Validating the Trace Analyzer for Memory Reuse Profiles}
We use reuse distance based analysis to estimate the data locality, thereby measure the cache performance. As these distances are independent of the hardware, we can use the same memory trace across all cache hierarchies and for different CPU models. Our data availability (final hit-rate, $P(h)$) experiments involve measuring the basic block reuse profiles ($P_i(d)$), final reuse profile ($P(d)$) and the conditional hit-rates ($P(h|d)$) at a given reuse distance $d$. We use $1\%$ sampling, where only $1\%$ of all the occurrences of a basic block are used in calculating the conditional reuse profiles of a benchmark. Note that the actual conditional hit-rates are measured using the same model shown in Eq.~\ref{eq:phd} with out actually applying any sampling as opposed to the predicted conditional hit-rates. To the best of our knowledge this is the best solution to measure the conditional hit-rates. Note that the over all hit-rates at different cache hierarchies can be measured using PAPI counters~\cite{mucci1999papi}, however, we used Byfl~\cite{Scott:Byfl} for this purpose, since Byfl uses the native LLVM instrumentation. Note that, we measure these reuse profiles for smaller inputs of a program (to avoid mammoth memory traces) and extrapolate (using the techniques similar to~\cite{ChennupatiSBTBM17}) them for larger inputs. These profiles are further used in estimating the hit-rates.

Figure~\ref{fig:sd_phit} shows the conditional hit-rates of {\em JACOBI} for all the three different cache hierarchies. In general, for better hit-rates of an application, area under the curve for hit-rates should have more reuse distances to the left of the waterfall-step in each of the graphs. The drop indicates that the reuse distance is beyond the size of the cache, which will not fit at that cache hierarchy, thus a cache miss inevitably results. In both optimized and unoptimized cases, $L_1$ has higher cache-misses than $L_2$ and $L_3$, trivially due to the increasing sizes of the respective cache hierarchies. 

\begin{table}[htp] 
	\centering
	\caption{Percentage of error between the predicted and actual cache hit-rates at different cache hierarchies when compiled with and without optimizations ({\em opt} and {\em Unopt})}\label{tab:hitrate} \vspace{-2ex}
	\begin{tabular}{|c|c|c|c|c|c|c|}
		\toprule
		\multirow{3}{*}{\textbf{Benchmark}} & \multicolumn{6}{c|}{\textbf{\% Error}}\\
		\cline{2-7}
		& \multicolumn{2}{c|}{\textbf{L$_1$}} & \multicolumn{2}{c|}{\textbf{L$_2$}} & \multicolumn{2}{c|}{\textbf{L$_3$}} \\
		\cline{2-7}
		& \textbf{Unopt} & \textbf{Opt}  & \textbf{Unopt} & \textbf{Opt} & \textbf{Unopt} & \textbf{Opt} \\
		\midrule
		{\em JACOBI} & 5.41 & 4.90 & 2.25 & 2.10 & 1.16 & 1.02\\
		{\em MATMUL} & 5.20 & 3.79 & 1.71 & 1.69 & 0.55 & 0.61 \\
		{\em LAPLACE2D} & 5.81 & 5.09 & 3.05 & 2.90 & 1.81 & 1.57 \\
		{\em BlackScholes} & 2.60 & 2.14 & 1.16 & 1.07 & 0.52 & 0.43\\
		\midrule
		\textbf{Average} & \textbf{4.75} & \textbf{3.98} & \textbf{2.04} & \textbf{1.94} & \textbf{1.01} & \textbf{0.98}\\
		\bottomrule
	\end{tabular}\vspace{-2ex}
\end{table}

\begin{figure*}[htp]
	\centering
	\includegraphics[width=1.0\linewidth]{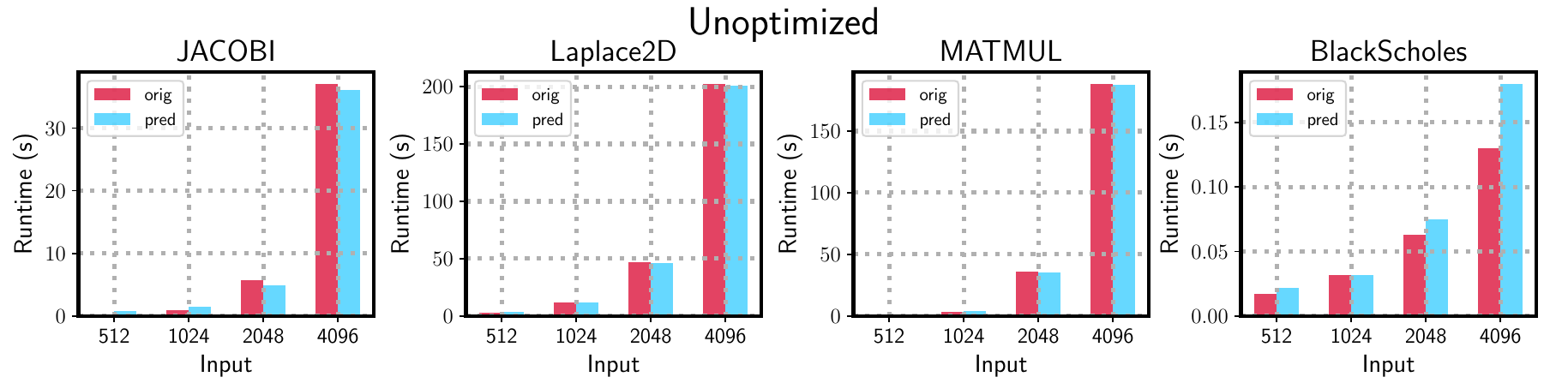}
	\includegraphics[width=1.0\linewidth]{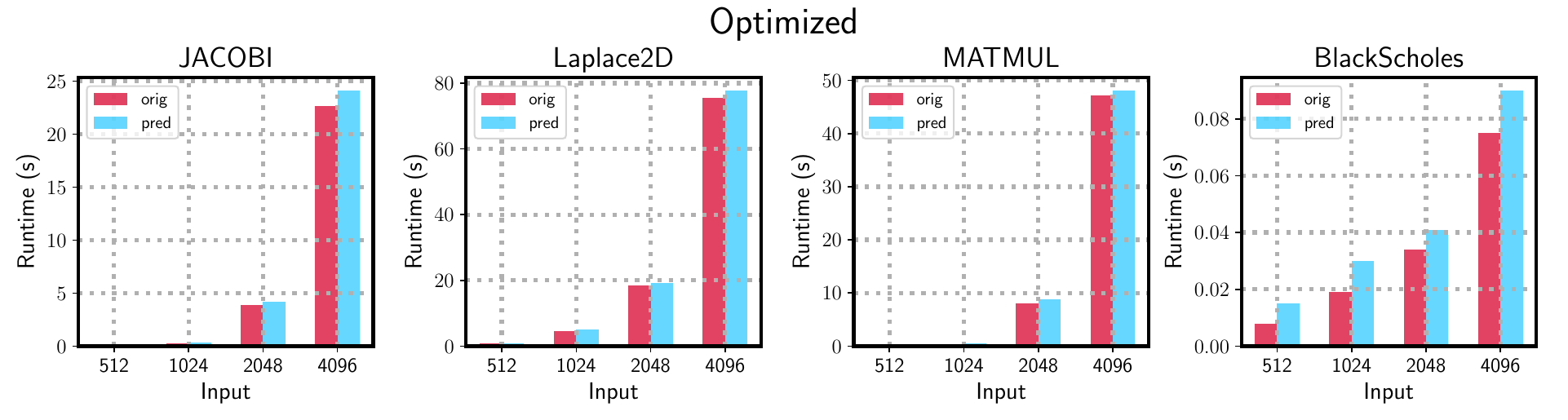}
	\vspace{-5ex}\caption{Actual versus predicted runtimes of all the four benchmarks with and without optimizations.}
	\label{fig:runtimes}
\end{figure*}

On the other hand, the results show a comparison between the predicted and the actual conditional hit-rates. In all the three cache hierarchies, the predicted results overlap with that of the actual hit-rates at a given reuse distance. Note that for $L_2$ and $L_3$ there is at most one reuse distance beyond the cache size, that stands for the reuse distance of $\infty$, resulted from the first time memory accesses of a program. However, this poses a problem for predictions with growing number of reuse distances for an increased input size.

In order to further strengthen the accuracy of the predicted cache hit-rates, we compare the percentage of error in the final hit-rates with and without optimizations for all the four benchmarks. Table~\ref{tab:hitrate} presents the percentage of error in our hit-rate predictions at different cache hierarchies when we compile the source code with and without optimizations. On an average, we have an error rate of $4.75\%$ on $L_1$ without optimizations, while that with optimizations is $3.98\%$. The lowest prediction error is reported to be $0.98\%$ on $L_3$ with optimizations. There is a slight difference in the hit-rate predictions between the optimized and unoptimized versions of the code. The difference is due to the fact that the optimized versions of the programs contain low locality of accesses in comparison with their unoptimized counterparts. In fact, in the optimized code, the compiler replaces some of the accesses to registers and constants. 

\begin{table}[htp]
	\centering
	\caption{Cache hit-rates for different input sizes of {\em JACOBI}}\label{tab:hits-input} \vspace{-2ex}
	\begin{tabular}{|c|c|c|c|c|}
		\toprule
		\multirow{2}{*}{\textbf{Cache}} & \multicolumn{4}{|c|}{\textbf{hit-rates}}\\
		\cline{2-5}
		& \textbf{512} & \textbf{1024} & \textbf{2048} & \textbf{4096} \\
		\midrule
		$L_1$ & 0.971 & 0.838 & 0.771 & 0.761\\
		$L_2$ & 0.997 & 0.924 & 0.816 & 0.785\\
		$L_3$ & 0.999 & 0.947 & 0.915 & 0.887 \\
		\bottomrule
	\end{tabular}\vspace{-2ex}
\end{table}
Our system offers the capability to predict the final hit-rates at different cache hierarchies. Table~\ref{tab:hits-input} shows the hit-rates for {\em JACOBI} at four different input sizes $512$, $1024$, $2048$ and $4096$. The predicted hit-rates indicate that the cache misses happen more often with the increase in the input size. Here, with the final hit-rates, we calculate the effective latency and throughput (see Eq.~\ref{eq:avg_latency}) for the memory accesses, thereby predict the runtimes of a program. 

We measure the time complexity of reuse profiles asymptotically. In general, the naive reuse profile calculation~\cite{Mattson:stack} has a computational complexity of $O$($NM$). Our analytical reuse profile calculation consumes a computational complexity of $O$($NSB$) $\sim$ $O$($N$), here the number of samples ($S$) and size of the basic block ($B$) are constant. The worst case complexity is $O$($NM$), when the {\em sampling rate} becomes $100\%$, which never happens due to the fixed sampling rate of $1\%$, which is sufficient to accurately approximate the reuse profiles. We can further optimize the complexity to $O(logN)$ using a parallel tree based implementation~\cite{niu2012parda}. Here, $N$ is the number of memory references in a trace.

\subsection{Validating the PPT Compute Node Simulator}

Bringing all these pieces together, we validate the predicted runtimes with that of the actual runtimes on Intel Xeon \texttt{E5-2695} at different input sizes for each of the benchmarks. The inputs vary from $512$, $1024$, $2048$, and $4096$ for each of the benchmarks. For the first three ({\em JACOBI}, {\em MATMUL} and {\em LAPLACE2D}) benchmarks, the respective input sizes represent both the dimensions of the matrix, while that for {\em BlackScholes} represents the dataset size. The pipeline model contains the following parameterized values.

Figure~\ref{fig:runtimes} compares the predicted runtimes with that of the actual for all the four benchmarks with and without optimizations. The X-axis represents the corresponding input sizes for each benchmark, while the Y-axis stands for runtime in seconds. The actual runtimes are recorded with the \textrm{C} codes executed on a single core using \texttt{taskset} utility. Clearly, the benchmarks compiled without optimizations take longer to execute when compared with that of the optimized. Overall, in both the cases of with and without optimizations, we are slightly over-predicting (except for {\em JACOBI} Unoptimized), nonetheless, the predictions are fairly accurate.

In the unoptimized case, the average percentage of error (across different inputs) between the predicted and actual runtimes for all the benchmarks is as follows: {\em JACOBI} has $10.58\%$, {\em LAPLACE2D} has $7.61\%$, {\em MATMUL} contains $12.48\%$ and {\em BlackScholes} shows $13.84\%$. In the optimized case, the average percentage of error results for the four benchmarks are $12.06\%$, $9.42\%$, $14.17\%$ and $15.01\%$ respectively. These accurate predictions are due to the fact that our model mimics the program execution in a real hardware through the combination of reuse profiles for data locality with the pipeline effects.

From the results, we observe that the runtimes are over predicted for {\em BlackScholes}, is due to the nature of the application. Contrary to the other three benchmarks, {\em BlackScholes} is compute bound rather memory bound. In this case, our extrapolations on the reuse profiles that mimic the memory reuse behavior contribute to the slight differences in the predictions. 

Another observation is that, when the input size is small, we see higher percentage of error in the prediction while that reduces drastically for larger inputs. For example, on {\em Laplace2D} for an input of $512$, the percentage of errors with and without optimizations are $8.35\%$ and $9.11\%$ respectively, which reduces significantly at the larger input size ($4096$) to $4.16\%$ and $1.55\%$ respectively. The reason for such an effect in the predictions is because of the cumulative effect of memory model and the pipeline execution. As the input sizes grow, the predictions of our model get much better.

\begin{table}[htp]
	\centering
	\caption{Basic block-wise generalized counts and the predicted runtimes for the kernel of the {\em JACOBI} (Opt)}\label{tab:bbtimes} \vspace{-2ex}
	\begin{tabular}{|c|c|c|c|c|c|}
		\toprule
		\multirow{1}{*}{\textbf{BB}} & \textbf{counts} & \textbf{runtime} & \textbf{counts $\times$ runtime}  \\
		\midrule
		1 & $n^2$ & 8.48e-8 & 1.423 \\
		2 & $n^2-n-1$ & 7.435e-8 & 1.247 \\
		3 & $n^2-2n+1$ & 8.56e-8 & 1.435 \\
		4 & $n^2-3n+2$ & 6.65e-7 & 11.155 \\
		5 & $n^2/2$ & 3.41e-8 & 0.285 \\
		6 & $k*n^2$ & 3.09e-8 & 5.704 \\
		7 & $k*n^2-n^2$ & 1.23e-8 & 1.857 \\
		8 & $k*n^2$ & 1.9e-9 & 0.318 \\
		9 & $k*n$ & 1.08e-8 & 0.0004 \\
		\midrule
		\textbf{Total}& \multicolumn{3}{|r|}{\textbf{23.427} seconds}\\
		\bottomrule
	\end{tabular} \vspace{-2ex}
\end{table}

Table~\ref{tab:bbtimes} shows the basic block-wise predicted runtimes for the kernel (contains $9$ basic blocks that influence the runtime, see Table~\ref{tab:graphlets}) of {\em JACOBI} compiled with optimizations. The generalized expressions are deduced from the machine learning model describe in section~\ref{sec:bbcounts}, the \textrm{runtime} column represents the time taken to execute a single iteration of a basic block and the last column shows the time taken for all the counts of a basic block. The results are for an input sizes of n=$4096$ and k=$10$, finally, runtime is expressed as a function of input, is devised as $4.51e^{-8}k*n^2+8.84e^{-7}n^2+1.08e^{-8}k*n-2.24e^{-6}n+1.34e^{-6}$. Similarly, we can express the runtime as a function of input size for other benchmarks.

Overall, PPT runtime predictions with AMMP are achieved in linear time in the number of lines of the input code ($C$), denoted as $|C|$ (independent of the asymptotic runtime of $C$). Pre-processing the input code takes $O(|C|(|I|))$ steps, where $|I|$ is the number of input parameters. For long running code, this simulation model runtime reduction is a big improvement in performance when compared to other instruction-level simulators, including traditional PPT. For codes with long running times, this runtime reduction in the simulation model is a ground-breaking improvement in performance when compared to other instruction-level simulators, including traditional PPT.

\section{Sensitivity Analysis}
\label{sec:sensitivity}
We illustrate how to use PPT-AMMP for the purpose of identifying performance resource bottlenecks or for exploring future architectures. We simply change some of the hardware parameters in $H$ that we feed to the the PPT Compute Node Simulator; in addition we also vary the $C$-code input parameter sizes $|I|$. Concretely, we modify pipeline counts, $L_1$ and $L_2$ cache sizes, and vary input sizes, with which we study the effect of these changes on runtimes. We start with the parameterized hardware model designed for Intel Xeon \texttt{E5-2695} as a baseline.

\subsubsection{Effect of Pipelines}
\begin{figure}[htp]
	\centering
	\begin{minipage}{0.33\textwidth}
		\centering
		\includegraphics[width=1.0\linewidth,height=0.85\linewidth]{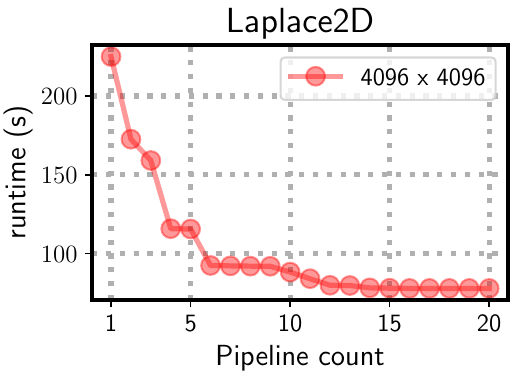}
		\vspace{-5ex}\caption{Effect of pipeline counts on runtimes}
		\label{fig:pipeline_counts} \vspace{-2ex}
	\end{minipage}\hfill
	\begin{minipage}{0.33\textwidth}
		\centering
		\includegraphics[width=1.0\linewidth,height=0.85\linewidth]{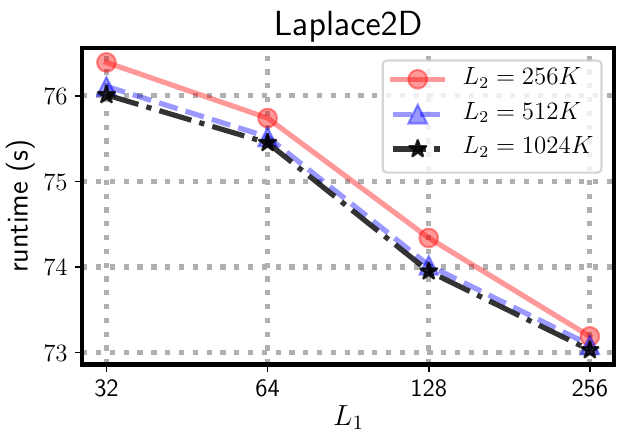}
		\vspace{-5ex}\caption{Effect of cache sizes on runtimes}
		\label{fig:l1_l2_sens} \vspace{-2ex}
	\end{minipage}
	\begin{minipage}{0.33\textwidth}
		\centering
		\includegraphics[width=1.0\linewidth,height=0.84\linewidth]{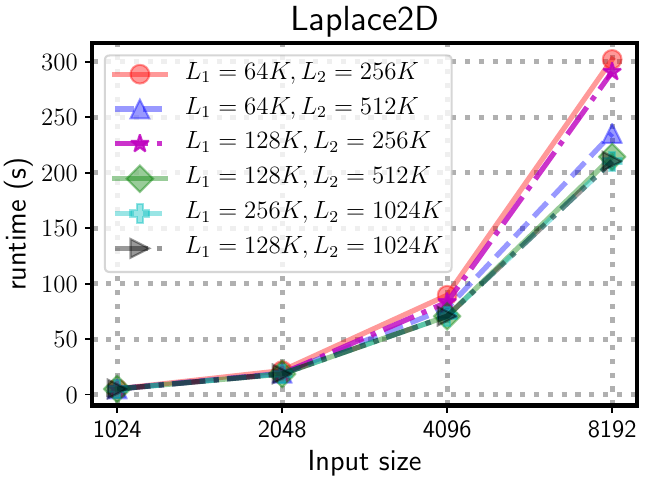}
		\vspace{-4ex}\caption{Effect of cache at different inputs}
		\label{fig:input_runtime} \vspace{-1ex}
	\end{minipage}
\end{figure}

Figure~\ref{fig:pipeline_counts} shows the effect of pipeline counts on runtimes for an example benchmark, {\em Laplace2D} (Opt) for an input mesh of size $4096 \times 4096$. We increase the dedicated number of pipelines for each of the CPU instructions (\textrm{iadd}, \textrm{fmul}, \textrm{fdiv}, etc). The results indicate that the runtimes are bigger for small number of pipeline dedicated pipelines. The runtimes keep decreasing as the number of pipelines increase, finally the runtime has no effect after a certain number of pipelines. This indicates that the pipelines are utilized efficiently in processing the instructions in a task-graphlet. Moreover, increasing the number of pipelines after a certain threshold has diminishing returns.

\subsubsection{Effect of Cache}
Figure~\ref{fig:l1_l2_sens} shows how runtime gets effected for a fixed input ({\em Laplace2D} with $4096$) with different $L_1$ sizes at a fixed $L_2$. We vary $L_1$ from $32$, $64$, $128$ and $256 K$ while fixing the $L_2$ at $256$, $512$ and $1024 K$. In these experiments, the runtimes decrease with the increase in cache sizes. When $L_2=256K$ the runtimes are expensive irrespective of $L_1$, on the other hand when $L_2$ increases, the runtimes are decrease and overlap. This indicate that having a relatively large $L_1$ and $L_2$ have positive impact on performance. We refrain to increase $L_1$ further due to the fact that we may see negative effects through increased latency in memory. 

\subsubsection{Inputs with Cache}
Figure~\ref{fig:input_runtime} shows how runtime changes with varying input work load and cache sizes. We vary input size of {\em Laplace2D} from $1024$, $2048$, $4096$ and $8192$ at different $L_1$ and $L_2$ configurations. The results indicate that for smaller input sizes, varying cache configurations has insignificant impact. At large input sizes ($4096$, and $8192$), $L_1=128K, L_2=512K$ has significant impact in reducing the runtimes. Nevertheless, increasing $L_1$ to $256$ and $L_2$ to $1024$ has insignificant impact on runtime. Therefore, we observe that the current input sizes fit in the hypothetical cache sizes of $L_1=128K$ and $L_2=512K$. Finally, from these two different (fixed and varying inputs) it is evident that increasing $L_1$ and $L_2$ up to a certain threshold has significant impact on performance.

\section{Case Study: SN (Discrete Ordinates) Application Proxy (SNAP)}
\label{sec:case-study}
In this section, we evaluate out approach, AMMP on a proxy, modern discrete ordinates neutral particle transport application, SNAP~\cite{snap}, which solves the radiation transport equation for
neutron and gamma transport. The resulting solution is the distribution
of these sub-atomic particles in space, direction of travel, particle
speed, and time. 

In SNAP, space is modeled in
three dimensions, \emph{x-y-z}. A finite volume discretization creates
a structured, Cartesian mesh of spatial mesh \emph{cells}.  SNAP implements the \emph{discrete ordinates} solution technique, where 
it computes the solution for a finite set of possible directions or
``angles.'' Each angle is associated with a particular weight, and
the solution for each angle may be performed individually. This scheme is
known as ``discrete ordinates''. The two degrees
of freedom manifest themselves in SNAP as a list of angles per octant in
3-D space and the eight octants themselves. Particle speed (or energy) is
binned into \emph{groups} that represent the sum of all particles in some
range of energy values. Lastly, the time dimension is discretized with a
finite difference solver. We have a system of equations in seven
dimensions (three in space, two in angle, one in energy, and one in time).

The governing transport equation is hyperbolic in the
space-angle dimensions as information flows from an upstream source to
downstream destinations. The solution for any given direction in the
discrete ordinates solution scheme requires the addition of particle
sources and sinks while stepping through
the spatial mesh cells. Colloquially it is known as a ``transport mesh
sweep.'' A mesh sweep can further be thought in terms of a task graph.
Such a graph for a mesh sweep on a structured mesh as we have in SNAP
is known a priori and is the same for all directions. This allows
scheduling and operation optimizations to be included explicitly in the
code.
This system of equations offers multiple levels of parallelism which are
exploited SNAP. The global spatial mesh is distributed across MPI ranks. In our study, we focus on a serial execution of SNAP only, leaving the study of various parallelism levels to future work.


SNAP accepts a number of input parameters, of which, the most prominent inputs are: {\em Nx} -- the number of uniformly-sized cells in the $\mathrm{X}$-direction; {\em Ny} -- the number of uniformly-sized cells in the $\mathrm{Y}$-direction; {\em Nz} -- the number of uniformly-sized cells in the $\mathrm{Z}$-direction; {\em Ichunk} -- the number of $\mathrm{X}$-planes for a single work chunk, ranges in (1, {\em Nx}); {\em Nmom} -- the order of the scattering expansion, ranges in (1, 4); {\em Nang} -- the number of discrete ordinates per octant, ranges in (1, 4); {\em Ng} -- the number of energy groups, ranges in (1, 4); {\em Li} -- the number of inner iterations per energy group, by default set to $5$, however, ranges in (1, 10); {\em Lo} -- the number of outer iterations per time step, ranges in (1, 10).

\begin{figure*}[htp]
	\centering
	\includegraphics[width=1.0\linewidth,height=0.35\linewidth]{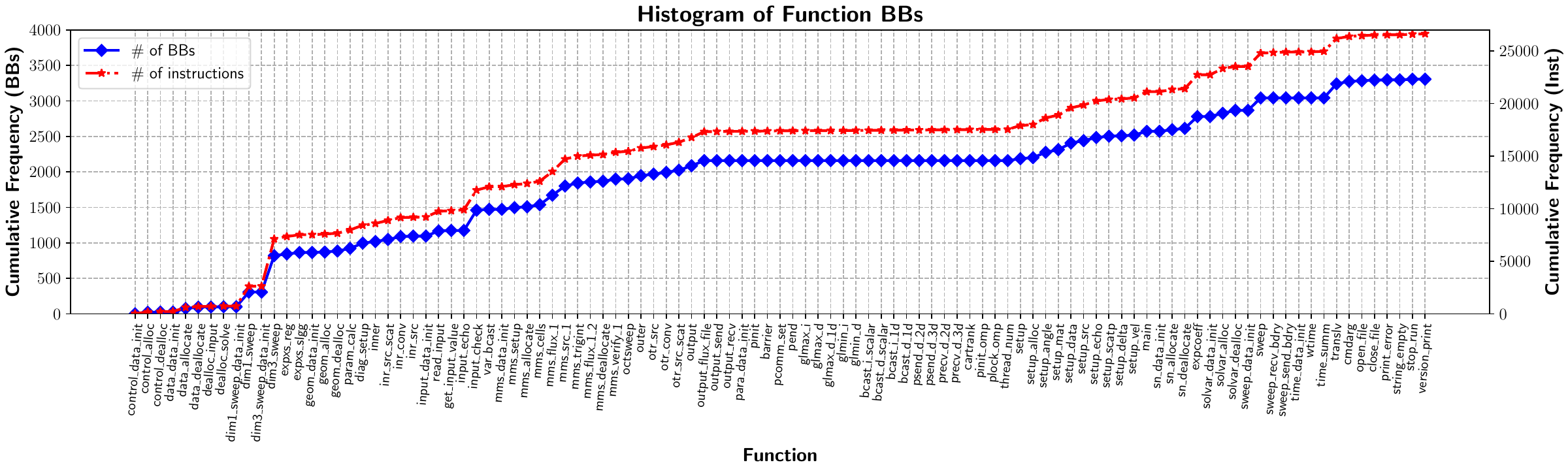}
	\caption{Cumulative frequencies of the basic blocks per function and the instructions for each of basic block of the functions.}
	\label{fig:bb_hist} \vspace{-2ex}
\end{figure*}

Figure~\ref{fig:bb_hist} shows the cumulative frequencies of the number of basic blocks in a function as well as the the number of instructions in each basic block of a function for SNAP. These histograms summarize the scale of the SNAP code. Overall, SNAP has $3306$ basic blocks spread across $103$ functions, of which $2669$ basic blocks are executed at least once while the rest are never executed. Note that those non-executing BBs stand for the error checking, termination, etc. Moreover, the number of times a basic block gets executed depends on the combination of input variables rather any one variable. These basic blocks in total contain $26638$ instructions. Of these basic blocks, {\em dim3\_sweep} contains a maximum number of basic blocks, $513$, is considered as the kernel of SNAP.

In this study, as opposed to employing a uni-variate model of basic block count prediction (see section~\ref{sec:bb_count_pred}), we use multi-variate machine learning models. In fact, the uni-variate models are not suitable for SNAP like applications given the complex nature of the multi-variate input combinations and their respective combinations. Therefore, we explore through two different prominent machine learning techniques in order to predict the basic block counts and the reuse profiles for different input sizes. The two machine learning techniques are \begin{enumerate*}[label=\itshape\alph*\upshape)] 
	\item symbolic regression and
	\item deep learning
\end{enumerate*} We describe the three approaches as follows.


\subsection{Symbolic Regression}
Symbolic Regression through Genetic Programming (GP)~\cite{Koza:gp}, is most commonly used technique to device functions that are in the unknown form, \(y = f(x_1, x_2, \dots, x_n)\) from multi-variate space using a finite number of samples. Here, \(x_1, x_2, \dots, x_n\) are independent variables, often called primitives in GP terminology. Inspired from the Darwinian theory of evolution, random initial population of solutions (functions) help to produce near-optimal functions for given data. In fact, these solutions have sex (crossover) and undergo through genetic mutations. As a result, qualitative solutions propagate to the next generation, this iterative process continues until convergence. A number of previous studies~\cite{vladislavleva2008order,chennupati2014predict,munoz2019evolving} show the success of symbolic regression through GP. 

\begin{figure}[htp]
	\centering
	\includegraphics[width=0.85\linewidth,height=0.35\linewidth]{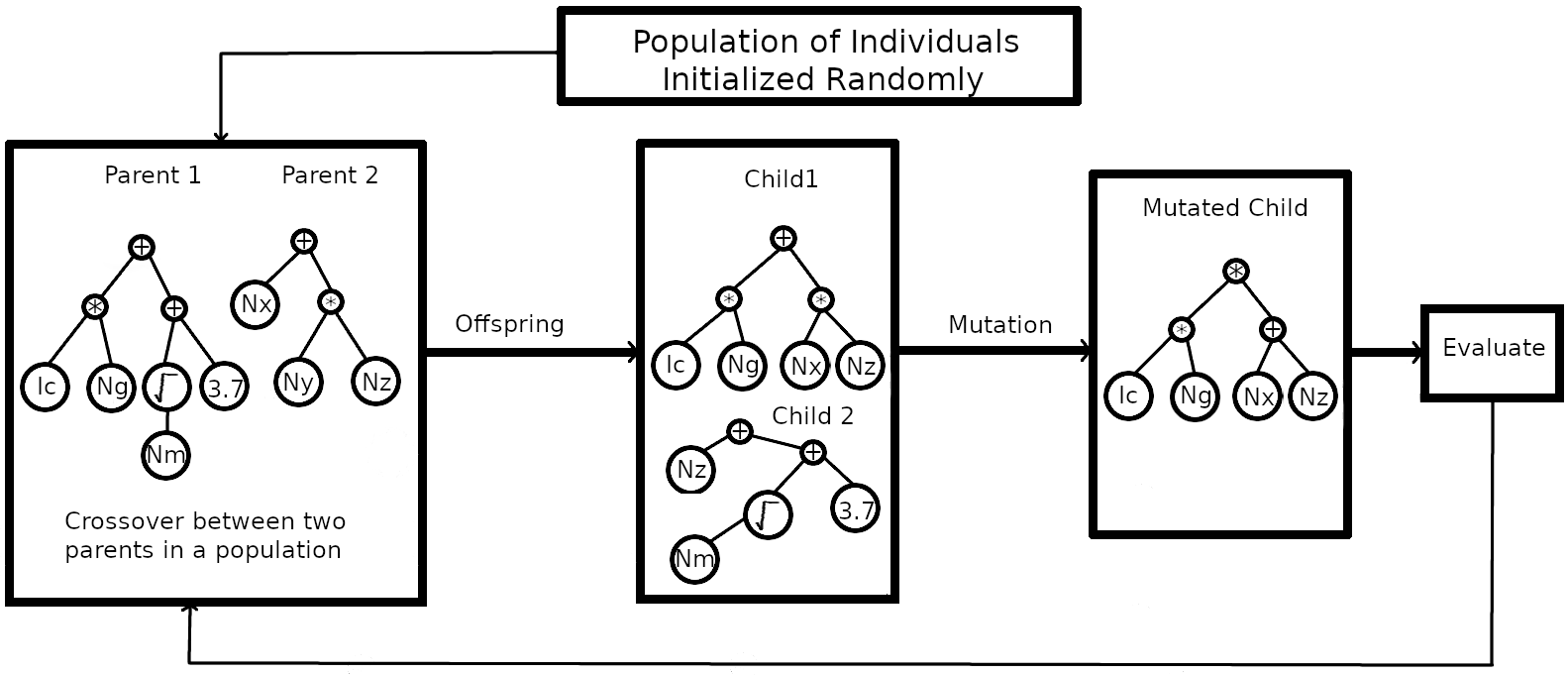}
	\caption{The genetic programming approach for symbolic regression to generate fits for basic block counts.}\label{fig:symreg}
\end{figure}

Figure~\ref{fig:symreg} shows the algorithmic cycle of genetic programming. The algorithm starts with a randomly initialized population of symbolic regression equations. These equations are represented as trees. The population undergoes a series of operations: i) crossover across pairs of symbolic regression equations; ii) mutations on the offspring from crossover; iii) the evaluation of the quality of the resultant programs and iv) the qualitative equations form the population for the next iteration. In symbolic regression for SNAP, GP takes the SNAP specific input variables ({\em Nx}, {\em Ny}, etc) along with the commonly used arithmetic operators specific to GP such as \( +, -, *, /, sqrt, cos, sin, etc\). For example, the function, \(Nx*Ny+Ichunk*Nz+sqrt(Nmom)+3.472\) derived from symbolic regression predicts the number of times the $9^{th}$ basic block (SNAP has a total of $3306$ basic blocks) gets executed. This prediction is general enough where the count changes with respect to the input size. Note that the form of the equation for this basic block is resulted from the genetic programming for symbolic regression. The hyper-parameters used in GP are as follows: the number of generations is $3000$, crossover rate of $0.9$ while the point mutation ratio of $0.01$, the maximum and minimum tree sizes are $75$ and $4$ and the respective tree depths are $15$ and $1$.

\subsection{Neural Arithmetic Logic Units}
Recently, deep Learning has become a de facto norm with the success in a number of applications. The dense neural networks, especially, with the dense and fully connected layers among neurons are the most common deep learning architectures. However, these dense networks are sub-optimal in predicting the numerical values. The recently introduced neural arithmetic logic unit (NALU)~\cite{trask2018neural}, addresses this inability of the dense networks to offer a systematic extrapolation of numerical values. In NALU, numerical values are represented as neurons. These neurons can undergo simple arithmetic operations such as \(+, -, *, /\). 

NALU contains two neural accumulators (NAC), first for addition/subtraction transformation and second for multiplication/division transformations. NAC is a special type of linear layer in a neural network. The transformation matrix (\textbf{W}) of first NAC layer contains \({-1, 0, 1}\) as the elements in order to guarantee additive/subtracting outputs using the inputs to the layer. Similarly, the second NAC layer (operates in {\em log} space) guarantees multiplicative/division transformations of the inputs. In fact, NALU can generalize better in predicting the unseen numeric values due to the existence of addition, subtraction, multiplication, division and power functions (for example, \(sqrt\)). We use the following hyper-parameters to train NALU: the number of steps are $300K$, learning rate is $1e-3$, the optimizer is {\em Adam} with betas $0.9$ and $0.99$, and the number of hidden layers are $32$. For more details on NALU, we refer the interested readers to~\cite{trask2018neural}.


\subsection{Model Evaluation}
Our evaluation of SNAP using AMMP is two fold: first, evaluate the predictive power of machine learning models; second, validate runtimes across multiple inputs. In  the first set of experiments, we device two models to predict basic block counts and the reuse profiles. These two models are used to predict the basic block counts and the reuse distances at an arbitrary input combinations of SNAP. Note that the resultant models are regressive in nature. 

The input dataset to device the regression models consists of $6480$ unique data points, which are resulted from different combinations of $9$ SNAP inputs (Nx, Ny, Nz, Ichunck, Nang, Nmom, Ng, Li and Lo). Memory traces at certain input combinations can be large, therefore, the SNAP inputs are selected such that the generated memory traces are small. These $6480$ data points are divided into two datasets (train and validation) with a split ratio of $80:20$. The training set is used to learn the regression models while the validation set is useful in measuring the accuracy of the trained model at regular intervals of the training process. Moreover, we prepare a test set with $834$ data points, which is used to perform the final predictions. Note that the input combinations of the test set produce slightly larger trace files. Nonetheless, such large trace generation can be avoided in future work through instrumentation techniques that allow dynamic sampling~\cite{Zhong:PLA,ChennupatiSBTBM17} of memory accesses.


\subsection{Predict the Number of Executions of a Basic Block}
\label{sec:bb_count_pred}
We validate both predictive models in terms of the accuracy of extrapolations, in terms of the actual number of executions of a given basic block for SNAP. Note that SNAP has a total of $2669$ active basic blocks, when compiled with \textrm{O3} optimizations. Of which, $1928$ basic block counts remain constant irrespective of the input to SNAP. Therefore, we perform predictions on the remaining $741$ basic blocks using both the models.

\begin{figure}[htp]
	\centering
	\begin{minipage}{0.49\textwidth}
		\centering
		\includegraphics[width=1.0\linewidth,height=0.85\linewidth]{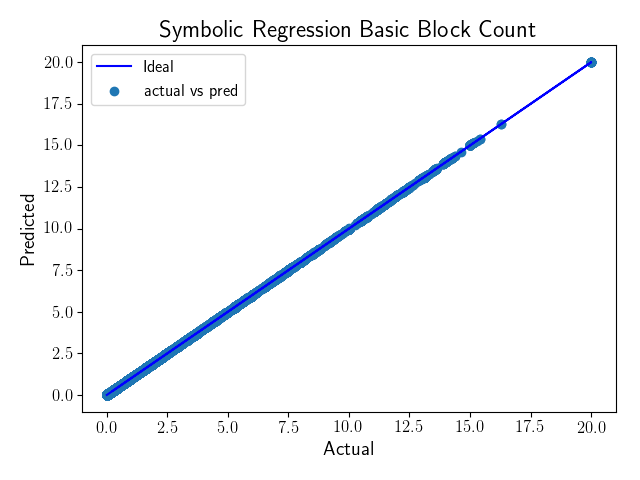}
		\vspace{-5ex}\caption{Symbolic regression extrapolation of basic block counts}
		\label{fig:reg_bb_count_extrapolate} \vspace{-2ex}
	\end{minipage}\hfill
	\begin{minipage}{0.49\textwidth}
		\centering
		\includegraphics[width=1.0\linewidth,height=0.85\linewidth]{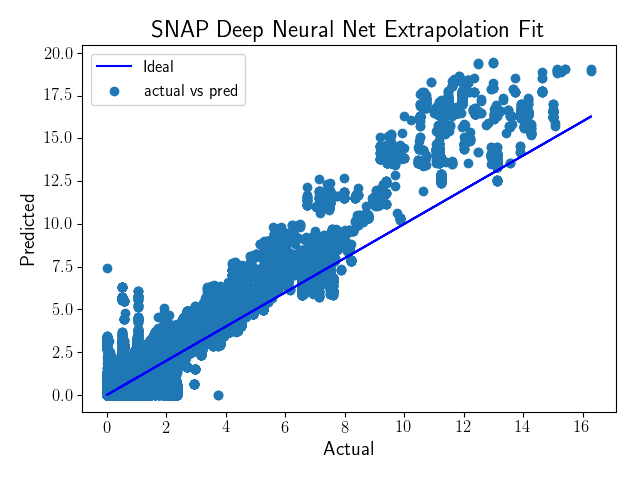}
		\vspace{-5ex}\caption{NALU extrapolation of basic block counts}
		\label{fig:dnn_bb_count_extrapolate} \vspace{-2ex}
	\end{minipage}
\end{figure}

Figure~\ref{fig:reg_bb_count_extrapolate} and~\ref{fig:dnn_bb_count_extrapolate} show the normalized extrapolations of actual versus predicted basic block counts for both the regression models. The predictions of the symbolic regression are better than that of the NALU predicted counts. This behavior is clearly evident in comparison with the ideal prediction (shown as a straight line). Although, the symbolic regression performs better than the deep learning, the predictions of latter are close to the ideal predictions. Symbolic regression performs better over NALU especially due to the fact that the equations of symbolic regression include arithmetic operators such as $sqrt$ as opposed to NALU. In other words, symbolic regression explores through the solutions that are unexplored by the deep learning algorithm, NALU. Moreover, the equations resulted from symbolic regression are interpretable as opposed to the black box nature of deep learning.

\begin{table}[htp]
	\centering
	\caption{The predicted equations of symbolic regression for a few functions and the respective basic blocks of SNAP}\label{tab:pred_eqs}
	\begin{tabular}{|l|l|l|}
		\toprule
		\textbf{Function} & \textbf{Basic block} & \textbf{Equation}  \\
		\midrule
		$dim3\_sweep$ & {\em for.inc115} & \begin{math}1.72 \times Nx\times Ny\times Nz\times Ng\times Lo\times(Li+0.23\times \sqrt{Li})\end{math} \\
		& {\em vector.body646} & \begin{math}0.073 \times Nx\times Ny\times Nz\times Ng\times Li\times Lo\times(Nmom-0.62/Nmom)\end{math}\\
		& {\em min.iters} & \begin{math}1.63 \times Nx\times Ny\times Nz\times Ng\times Li\times(Lo+0.317\times \sqrt{Lo})\end{math} \\
		\hline
		$mms\_flux\_1$ & {\em vector.body4720} & \begin{math}0.20 \times Nz \times Ng\times Lo\times(2.49-0.001\times Lo\times \sqrt{Nz}) \end{math} \\
		& {\em land.lhs} & \begin{math}0.21 \times Nz\times Ng\times Lo\times (2.38-0.0004\times Nz\times Lo) \end{math} \\
		& {\em if.then272} & \begin{math} 0.25 \times Ny \times Ng \times Lo \times(Nz+0.086/Lo) \end{math} \\
		\hline
		$translv$ & {\em for.cond354} & \begin{math} 0.221 \times Nx \times Ny \times Nz \times Ng \times Lo\times(Li+0.184\times \sqrt{Li}) \end{math} \\
		& {\em for.inc425} & \begin{math} 0.411 \times Ng \times Li \times Lo \times(2.452-0.004\times Li\times Lo)  \end{math} \\
		& {\em middle.block1195} & \begin{math} 0.484 \times Nz \times Ng \times Lo \times (Li+(-0.204)/((Li-3.854 \times Ng))) \end{math} \\
		& {\em if.end103} & \begin{math} 0.0862 \times Nx \times Ny \times Nz \times Ng \times Li \times Lo \times (Nmom-0.484/(Nmom))\end{math} \\
		\bottomrule
	\end{tabular}
	
\end{table}

Table~\ref{tab:pred_eqs} presents a few regression models devised using symbolic regression. The equations are general enough to fit an  arbitrary input combination of SNAP. On the other-hand, the deep neural net trained model is black-box in nature, which poses a difficult task in interpretation. The symbolic regressions produced equations concise and easy to interpret while satisfying the first principles of hardware architecture. Considering the accuracy of predictions and the interpretability we use the predictions of symbolic regression in PPT to predict the runtimes of SNAP at various inputs.

\subsection{Reuse Profile Predictions}
\label{sec:bb_rd_pred}
We extrapolate the reuse distances. In order to extrapolate the reuse distances, we use the memory traces produced for the above $6480$ inputs (same inputs used in predicting the basic block counts) of SNAP. As opposed to the single-variate reuse distance prediction in~\cite{ChennupatiSBTBM17}, ours is multi-variate prediction. Our predictions are based on the above mentioned input variables for SNAP while the output predictions are based on average reuse distances.

An important challenge with reuse distances is that the number of reuse distances change with the increase in the input sizes. Predicting these varying distances across $6480$ data points. Therefore, we apply a binning strategy thereby predict a fixed number of average reuse distances. An interesting observation with reuse distances is that a few number of initial reuse distances are consistent irrespective of the input size. We ignore these consistent reuse distances from prediction while predicting the average reuse distance of the bins. In our strategy, we bin the changing reuse distances for each of the data point into $40$ bins, thus will have $40$ regression models.

\begin{figure}[htp]
	\centering
	\begin{minipage}{0.49\textwidth}
		\centering
		\includegraphics[width=1.0\linewidth,height=0.85\linewidth]{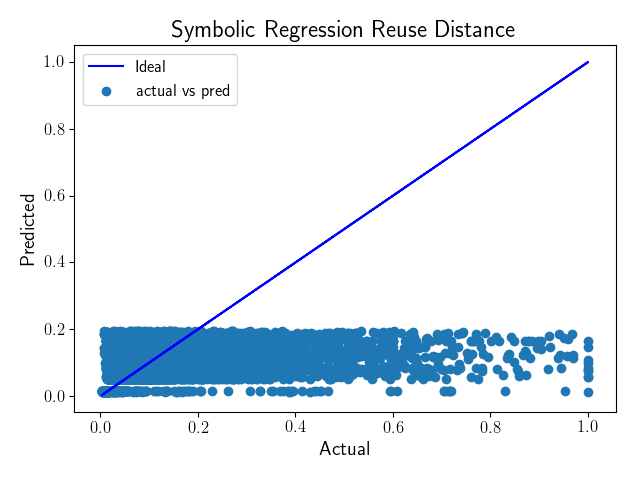}
		\vspace{-5ex}\caption{Symbolic regression extrapolation of reuse distances}
		\label{fig:reg_sd_extrapolate} \vspace{-2ex}
	\end{minipage}\hfill
	\begin{minipage}{0.49\textwidth}
		\centering
		\includegraphics[width=1.0\linewidth,height=0.85\linewidth]{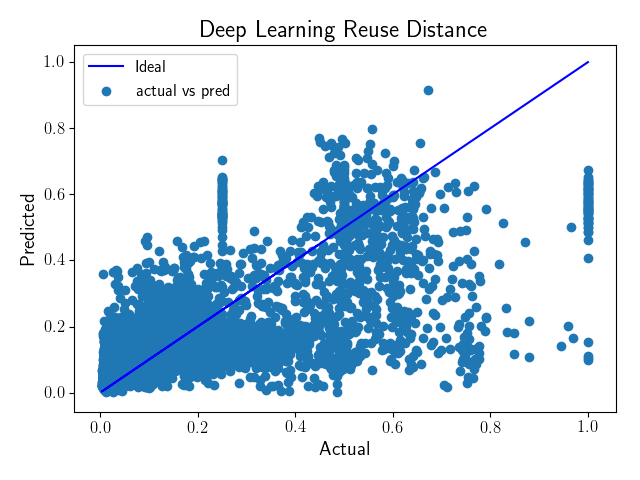}
		\vspace{-5ex}\caption{NALU extrapolation of reuse distances}
		\label{fig:dnn_sd_extrapolate} \vspace{-2ex}
	\end{minipage}
\end{figure}

Figure~\ref{fig:reg_sd_extrapolate} and~\ref{fig:dnn_sd_extrapolate} show the normalized extrapolated reuse distances with respect to that of the actual distances for both symbolic regression and deep neural networks. Symbolic regression extrapolations for reuse distance are under predicted while that of deep learning has less error in prediction. Although the deep learning model is hard to interpret, the performance of the predictions for reuse distance are accurate. Therefore, we decided to use the predicted reuse distances from deep neural networks as inputs to the AMMP.

\subsection{Discussion}
We discuss the a number of lessons learned during the regression model building. We overcome several complications in producing our models especially the symbolic regression. First, we excluded a few basic blocks from prediction as the corresponding counts are constant irrespective of the input size. Second, the indicator variables (such as Nmom, Ichunk, Nang, Ng) have a distinct impact on the performance due to the converging behavior of SNAP. Thus, the regressions for these basic blocks that rely on these indicator variables must be performed separately. In general, ignoring the indicator variables during model building helps to deduce near optimal solutions with symbolic regression and deep learning. Another unforeseen problem, is to in determine the forms of equations that performed better on extrapolations when identical performance of different forms were attained in the training and test data sets. For example, typically, better not use symbolic regression with the inclusion of square roots when an alternative and equally performing solution is available. In the case of reuse distance, the under-prediction of symbolic regression and the near feasible predictions (not optimal) of deep neural networks is because of the fact that we are predicting the average values rather than the absolute reuse distances. Moreover, due to the robustness of the memory model (discussed in section~\ref{sec:traceanalyzer}), the effects of these predictions is minimal on hit-rate. This is due to the fact that every reuse distance beyond the given cache size will be a miss and our model includes that effect. The average bin-wise reuse distances help us identify the reuse distance that range within the cache sizes. Hence, we find that our regression models in bot h the cases of basic block counts and the reuse distances are efficient in task of prediction.

On the other hand, modern processors have multi-cores with both shared and private caches. Unlike sequential programs, the reuse profile calculation of a parallel program is architecture-specific. The thread of a core accesses the private cache while all the cores access the shared cache. The works in~\cite{Jiang:RD-Applicable-on-chip} model the shared and private caches using the reuse profiles. Recently researchers proposed an analytical model and sampling to speed up the performance prediction~\cite{Jiang:RD-Applicable-on-chip,Schuff:2010:AMR:1854273.1854286,Multicore-Aware-Derek}. All these models require collection of large traces from parallel execution of an application from threads. Contrarily, we can collect trace once from the sequential run of the application, predict shared cache performance for different number of threads using interleaving strategies (similar to~\cite{atanu:sasmm}) in order to mimic the parallel execution of a program. This helps to generate a scalable model. Reuse distance analysis for multi-cores requires specific focus, therefore, we reserve our attempts for future work.

\subsection{Runtime Predictions}
We validate the actual runtimes with that of the predicted runtimes using PPT~\cite{ppt} toolkit. The inputs to the PPT-AMMP stylized application model of SNAP are tow fold: software and hardware parameters. The software input parameters are predicted basic block counts, predicted reuse profiles and the task graphlets, whereas the hardware parameters are instruction level pipeline latencies and the corresponding throughput of a target architecture (in our case, \texttt{Intel Xeon E5-2695} processor with a clock speed of $2.1$ GHz). The simulated processor uses the following parameters: $L_1$ data cache latency of $4$ cycles, $L_2$, $L_3$ and RAM latencies of $12$, $43$ and $65$ cycles respectively. The associativity of the three caches are $8$, $8$ nd $20$ respectively, while the cache line sizes for all the three caches are $64$. The maximum number of threads $36$, while the physical cores are $18$, however, our application uses single thread as the application relies on sequential execution.

\begin{table}[htp]
	\centering
	\caption{Actual versus predicted runtimes of SNAP}\label{tab:snap_times_compare}
	\begin{tabular}{|c|c|c|c|c|c|c|c|c|c|c|c|c|c|}
		\toprule
		\multirow{2}{*}{\textbf{\#}} & \multicolumn{9}{|c|}{\textbf{Input}} & \multirow{2}{*}{\textbf{converged?}} & \multicolumn{2}{|c|}{\textbf{Runtime (s)}} \\
		\cline{2-10}\cline{12-13}
		& \textbf{Nx} & \textbf{Ny} & \textbf{Nz} & \textbf{Ichunk} & \textbf{Nmom} & \textbf{Nang} & \textbf{Ng} & \textbf{Li} & \textbf{Lo} & & \textbf{Actual} & \textbf{Predicted} \\
		\midrule
		in1 & 32 & 40 & 48 & 1 & 4 & 80 & 5 & 5 & 100 & yes & 51.58 & 58.02\\
		in2 & 32 & 20 & 48 & 1 & 4 & 80 & 5 & 5 & 100 & yes & 151.92 & 156.74 \\
		in3 & 32 & 48 & 20 & 1 & 4 & 2 & 1 & 6 & 7 & yes & 5.012 & 5.901 \\
		in4 & 10 & 20 & 48 & 10 & 3 &  2  & 4  & 6 &  500 & yes & 20.66 & 20.31\\
		
		\bottomrule  
	\end{tabular}
\end{table}

Table~\ref{tab:snap_times_compare} compares the actual runtimes of SNAP with that of the predicted runtimes for four different input sizes. The results indicate that our predictions are accurate irrespective of the convergence of SNAP. However, the runtimes are over-predicted, this is due the fact that the predictions rely on the pipeline latencies and throughput of target architecture pipelines. Accurate micro bench-marking to measure the instruction latencies is crucial and is out of the scope of this paper. Finally, we summarize that the proposed model with multi-variate predictions is accurate and efficient in predicting the performance of an application like SNAP. Moreover, our model does not take into effect the speculative execution. These two factors contribute to the over prediction, focus of the future research.

\subsection{Runtime Predictions -- Different Architectures}
In order to test the generality of our proposed approach and the resultant predictions, we consider four more architectures. These four architectures belong to different generations of Intel series of processor micro-architectures: {\em Ivy Bridge}, {\em Haswell}, {\em Broadwell} and {\em Skylake}. Note that the selection of these architectures is purely based on our capability to access the corresponding CPUs to run SNAP in real time and collect the actual runtimes. Table~\ref{tab:more_arch} shows the hardware parameters used in simulating the respective architecture.

\begin{table}[htp]
	\centering
	\caption{Validating the runtime predictions on multiple current hardware architectures}\label{tab:more_arch}
	\begin{tabular}{c|c|c|c|c}
		\toprule
		\toprule
		\multirow{3}{*}{\textbf{Hardware}} & \textbf{Parameters} & \multirow{3}{*}{\textbf{Input}} & \multicolumn{2}{c}{\multirow{2}{*}{\textbf{Runtime (s)}}} \\
		& cache, sizes, cache line size, associativity, & \\
		\cline{4-5}
		& cache \& RAM latency, threads, clock speed (Hz) & & \textbf{Actual}  & \textbf{Predicted} \\
		\midrule
		\midrule
		\multirow{4}{*}{Ivy Bridge} & & in1 & 9.164 & 8.99\\
		& 3-level cache, (64 KB, 256 KB, 20 MB), (64, 64, 64), (8, 8, 12), & in2 & 139.91 & 145.53\\
		& (4, 12, 30) \& 36.72 cycles, 16 threads/8 cores, 2.60 GHz & in3 & 2.86 & 2.15\\
		& \texttt{Intel E5-2650} & in4 & 17.50 & 18.01\\
		\hline
		\hline
		\multirow{4}{*}{Haswell} & & in1 & 7.318 & 6.92\\
		& 3-level cache, (32 KB, 256 KB, 25 MB), (64, 64, 64), (8, 8, 16), & in2 & 136.54 & 143.38\\
		& (4, 12, 43) \& 62 cycles, 40 threads/20 cores, 3.10 GHz & in3 & 2.27 & 2.15\\
		& \texttt{Intel E5-2687W} & in4 & 14.64 & 15.33\\
		\hline
		\hline
		\multirow{4}{*}{Broadwell} & & in1 & 8.215 & 7.45\\
		& 3-level cache, (64 KB, 256 KB, 20 MB), (64, 64, 64), (8, 8, 20), & in2 & 139.88 & 142.43\\
		& 4, 12, 65) \& 38 cycles, 64 threads/32 cores, 2.10 GHz & in3 & 2.46 & 2.13\\
		&  \texttt{Intel E5-2683}  & in4 & 16.72 & 15.88\\
		\hline
		\hline
		\multirow{4}{*}{Skylake} & & in1 & 6.812 & 7.16\\
		& 3-level cache, (64 KB, 1 MB, 19.25 MB), (64, 64, 64), (8, 8, 20), & in2 & 63.31 & 67.10\\
		& (4, 14, 38) \& 42 cycles, 24 threads/12 cores, 3.40 GHz & in3 & 2.03 & 1.85\\
		& \texttt{Intel Gold 6138} & in4 & 13.33 & 14.68\\
		\bottomrule
		\bottomrule
	\end{tabular}
\end{table}

Table~\ref{tab:more_arch} compares the predicted runtimes with that of the actual runtimes for four different input variations presented in Table~\ref{tab:snap_times_compare} for SNAP across all the architectures. We observe that the predicted runtimes are fairly close to that of the actual. On an average, the percentage of error in runtime predictions are as follows: $8.38\%$ for Ivy Bridge, $6.54\%$ for Haswell, $7.37\%$ on Broadwell and $7.54\%$ on Skylake. These average error rates are below $10\%$ for all the predictions across different architectures, which signifies the strength of AMMP in generalizing the predictions across multiple architectures. Note that the program can be run on one of these platforms and the reuse profiles (because these are architecture independent) can be reused for other architectures.

\begin{table}[htp]
	\centering
	\caption{Time taken by PPT-AMMP to make the predict the performance of SNAP on four inputs across four different architectures}\label{tab:ppt_runtimes}
	\begin{tabular}{c|c|c|c|c}
		\toprule
		\toprule
		\multirow{2}{*}{\textbf{Parameters}} & \multicolumn{4}{c}{\textbf{Runtime on hardware model}}\\
		\cline{2-5}
		& \textbf{Ivy Bridge} & \textbf{Haswell} & \textbf{Broadwell} & \textbf{Skylake} \\
		\midrule
		\midrule
		in1 & 1.244 & 1.123 & 1.175 & 1.165 \\
		in2 & 1.251 & 1.253 & 1.212 & 1.223 \\
		in3 & 1.269 & 1.253 & 1.247 & 1.236 \\
		in4 & 1.247 & 1.226 & 1.140 & 1.205 \\
		\midrule
		\midrule
		Average & 1.252 & 1.214 & 1.194 & 1.207\\
		\bottomrule
		\bottomrule
	\end{tabular}
\end{table}

Table~\ref{tab:ppt_runtimes} shows the time taken by PPT-AMMP to make the predictions at a given input for all the four architectures. On average, PPT-AMMP consumes similar time to make the predictions irrespective of the input size as well as the simulated architecture under execution. This small overhead by the simulator is negligible compared to the performance predictions on various architectures without having to rerun the application in real time.

\section{Conclusion}
\label{sec:conclusions}
We showed the Analytical Memory Model with Pipelines (AMMP) for Performance Prediction Toolkit (PPT). The goal of the AMMP approach is to predict the performance of a software on a target architecture. PPT-AMMP helps to explore (in a quick manner) the software-hardware design space in order to find the best target architectures for the software, thereby recommends the future architectures. PPT-AMMP can be used to test various algorithmic variations on a number of target architectures. These capabilities provide insights about both software and hardware, which help in future hardware purchases as well as testing algorithmic variations before actual implementation in real time. 

Given the high-level source code, parameterized hardware and software parameters PPT-AMMP predicts the runtime. It offers accurate, scalable and reliable predictions due to the use of Markovian-style model of actual code combined with analytical modeling. AMMP prediction involves a number intermediate results, namely the data availability (reuse profiles and cache hit-rates) for the processor, latency and throughput of memory accesses and finally, the runtimes. We validated our model on four standard computational physics benchmark codes with and without optimizations. The average rate of error in prediction results align between $8\%$ (without optimizations) and $14\%$ (with optimizations), while the maximum error showed is $15.01\%$. We then studied the impact of changing the number of pipelines in a model and the cache sizes. We observe that the runtimes increase as the number of pipelines decrease, similarly increasing the cache sizes has a reciprocal effect on runtime. In future work, we plan to integrate the AMMP part of PPT with the MPI model for scalable predictions on large scale clusters; we will extend our reuse distance analysis with pipelines on GPUs, and we plan to more systematically study for what types of input codes, the PPT-AMMP works well. It is clear that for certain classes of codes, AMMP will be challenged to produce accurate results, for instance for codes that have instance-dependent convergence properties that do not exclusively rely on size. And as a final concession to theory: performance prediction of this sort cannot be guaranteed to work for any input code as this would amount to solving the (undecidable) halting problem.

We further envision to extend the proposed approach to multi-threaded applications as well as the GPU based programs. In fact, our recent implementation~\cite{atanu:sasmm} mimics the memory reuse profiles for multiple cores with shared and private caches. We further propose to extend this approach intertwined with the pipeline modeling of AMMP for more accurate predictions on different hardware architectures. Considering the performance modeling of GPU applications, we further extend our previous attempts~\cite{ArafaBCBSE20} to the distributed GPU algorithms (the modern neural networks use distributed GPUs) and different interconnects such as NVLink.
\bibliographystyle{ACM-Reference-Format}


\end{document}